\documentclass{article}
\usepackage{microtype}
\usepackage{graphicx}
\usepackage{booktabs} 
\usepackage{hyperref}
\usepackage{bm}
\usepackage{float}
\usepackage{multirow}
\usepackage{colortbl}
\definecolor{colorh}{rgb}{1,0.60,0.20}
\definecolor{colorm}{rgb}{1,0.72,0.30}
\definecolor{colorl}{rgb}{1,0.88,0.70}

\usepackage[accepted]{main}

\usepackage{amsmath}
\usepackage{amssymb}
\usepackage{mathtools}
\usepackage{amsthm}
\usepackage{wrapfig}
\usepackage{subfig}
\usepackage[capitalize,noabbrev]{cleveref}
\theoremstyle{plain}
\newtheorem{theorem}{Theorem}[section]

\theoremstyle{definition}
\newtheorem{definition}[theorem]{Definition}

\theoremstyle{remark}

\usepackage[textsize=tiny]{todonotes}

\icmltitlerunning{MLICv2: Enhanced Multi-Reference Entropy Modeling for Learned Image Compression}
\begin{document}

\onecolumn
\icmltitle{MLICv2: Enhanced Multi-Reference Entropy Modeling for Learned Image Compression}



\icmlsetsymbol{equal}{*}

\begin{icmlauthorlist}
\icmlauthor{Wei Jiang}{i1}
\icmlauthor{Yongqi Zhai}{i1}
\icmlauthor{Jiayu Yang}{i2}
\icmlauthor{Feng Gao}{i3}
\icmlauthor{Ronggang Wang}{i1,i2}
\end{icmlauthorlist}

\icmlaffiliation{i1}{Guangdong Provincial Key Laboratory of Ultra High Definition Immersive Media Technology, Shenzhen Graduate School, Peking University}
\icmlaffiliation{i2}{Pengcheng Laboratory}
\icmlaffiliation{i3}{School of Arts, Peking University}

\icmlcorrespondingauthor{Wei Jiang}{wei.jiang1999@outlook.com}
\icmlcorrespondingauthor{Ronggang Wang}{rgwang@pkusz.edu.cn}

\icmlkeywords{Machine Learning, ICML}

\vskip 0.3in



\printAffiliationsAndNotice{} 
\begin{abstract}
  Recent advances in learned image compression (LIC) have achieved remarkable performance improvements 
  over traditional codecs. Notably, 
  the MLIC series—LICs equipped with multi-reference entropy models—have substantially surpassed conventional 
  image codecs such as Versatile Video Coding (VVC) Intra. However, existing MLIC variants suffer from several limitations:
  performance degradation at high {bit-rate}s due to insufficient transform capacity, suboptimal entropy modeling that fails 
  to capture global correlations in initial slices, and lack of adaptive channel importance modeling. 
  In this paper, we propose MLICv2 and MLICv2$^+$, 
  enhanced successors that systematically address these limitations through improved transform design, 
  advanced entropy modeling, and exploration of the potential of instance-specific optimization.
  For transform enhancement, we introduce a lightweight token mixing block inspired by the MetaFormer architecture, 
  which effectively mitigates high-{bit-rate} performance degradation while maintaining computational efficiency. 
  For entropy modeling improvements, we propose hyperprior-guided global correlation prediction to extract global context 
  even in the initial slice of latent representation, complemented by a channel reweighting module that dynamically emphasizes 
  informative channels. 
  We further explore enhanced positional embedding and guided selective compression strategies for superior context modeling. 
  Additionally, we apply the Stochastic Gumbel Annealing (SGA) to demonstrate the potential for further performance improvements through input-specific optimization.
  Extensive experiments demonstrate that MLICv2 and MLICv2$^+$ achieve state-of-the-art results, reducing Bjøntegaard-Delta Rate by 16.54\%, 21.61\%, 16.05\% and 20.46\%, 24.35\%, 19.14\% 
  on Kodak, Tecnick, and CLIC Pro Val datasets, respectively, compared to VTM-17.0 Intra.
\end{abstract}
\section{Introduction}
  \label{sec:intro}
    Image compression is a fundamental problem in multimedia communication, aiming to achieve compact representation while maintaining high reconstruction quality. 
Traditional codecs such as VVC Intra~\cite{bross2021vvc} rely on block-based hybrid coding frameworks refined over decades. 
However, as these hand-crafted methods approach performance saturation, 
learned image compression (LIC) has emerged as a new paradigm that leverages 
end-to-end optimization to outperform conventional codecs.\par
{Recent LIC approaches~\cite{balle2016end,theis2017lossy,balle2018variational,minnen2018joint,minnen2020channel,
cheng2020learned,xie2021enhanced,he2021checkerboard,he2022elic,zou2022the,jin2025neural,zhao2023universal,tu2024reconstruction,
zhu2022transformerbased,koyuncu2022contextformer,duan2023qarv,wang2022neural,guo2021causal,wu22block,yang2024perceptual,
mishra2021multi,mishra2020wavelet,mishra2022deep2,mishra2022deep,park2023scalable,
chen2021nic,liu2023learned,jiang2022mlic,jiang2023mlicpp,jiang2024llic} 
typically follow an autoencoder-based framework optimized for the rate-distortion trade-off.
An analysis transform encodes the image into latent representations, which are quantized and decoded by a synthesis transform. 
Entropy models then estimate the probability distribution of quantized latents to enable efficient coding. 
With advanced transform designs and entropy models, several LIC systems~\cite{he2022elic,jiang2022mlic,jiang2023mlicpp,jiang2024llic,liu2023learned,li2024frequencyaware,lu2025learned,feng2025linear} 
have achieved or even surpassed the performance of state-of-the-art traditional codecs such as VVC Intra~\cite{bross2021vvc}.}\par
Entropy modeling is crucial to LIC performance as it minimizes conditional entropy by capturing informative priors. 
Hyperprior models~\cite{balle2018variational} introduce side information, while context models~\cite{minnen2018image,he2021checkerboard} 
capture spatial dependencies among neighboring elements. 
{Building upon these foundations, Jiang \textit{et al.} proposed the MLIC series~\cite{jiang2022mlic,jiang2023mlicpp}, 
\begin{figure}
  \centering
  \includegraphics[width=0.5\linewidth]
  {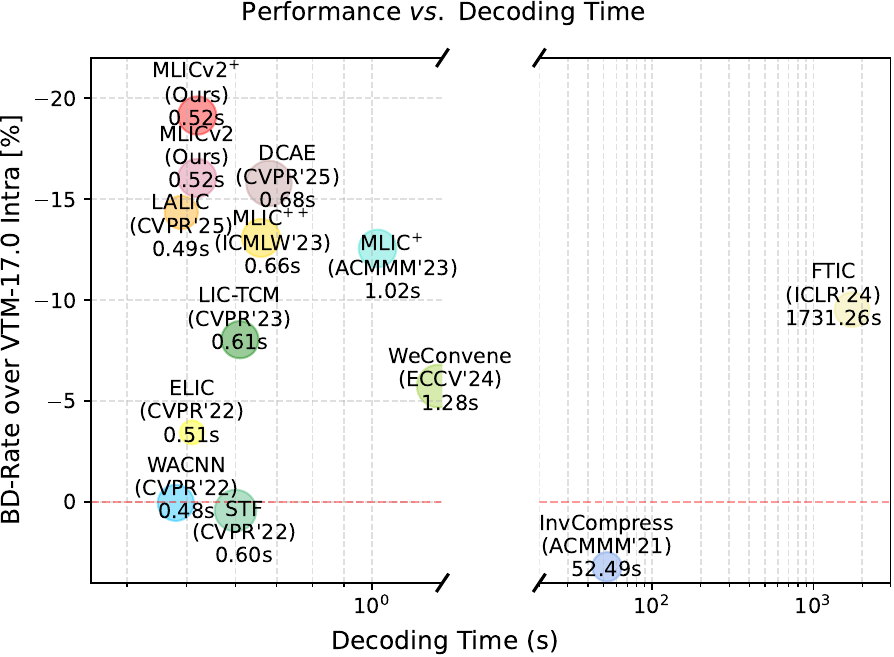}
  \caption{{BD-Rate and decoding time during inference on CLIC Professional Valid~\cite{CLIC2020} with 2K resolution.
  The bubble size indicates the number of model parameters.}}
  \label{fig:bdbr}
  \end{figure} 
which partitions latents into channel-wise slices~\cite{minnen2020channel} to model both inter- and intra-slice 
{local\footnote[1]{{“Local” denotes neighboring regions, specifically within the receptive field of a convolution or a windowed attention.}} 
and global\footnote[2]{{“Global” refers to the entire feature map.}}} dependencies. 
Within each slice, checkerboard partitioning~\cite{he2021checkerboard} enables two-pass decoding, and attention modules model global 
correlations from previous slices. 
To reduce computational cost, MLIC$^{++}$~\cite{jiang2023mlicpp} replaces standard attention~\cite{vas2017attention} 
with linear attention~\cite{shen2021efficient}\footnote[3]{{``Linear" denotes complexity linear with spatial resolution.}}, 
which scales more efficiently while maintaining global correlation modeling.}\par
{Despite their strong performance, MLIC models still face several limitations that motivate this work. 
First, in terms of transform design, performance degrades at high bit-rates due to limited transform capacity, which restricts the preservation of latent information. 
Second, for entropy modeling, global correlation estimation relies solely on previously decoded slices, preventing the capture of global dependencies in early slices. 
Furthermore, all channels are treated equally without considering their varying informational importance, potentially reducing modeling accuracy.}
\par
To systematically address these limitations, {and as illustrated in Fig~\ref{fig:change}}, we propose MLICv2, featuring enhanced transform and entropy modeling modules, and MLICv2$^+$, 
which additionally incorporates iterative refinement techniques~\cite{yang2020improving} to demonstrate framework potential.
\par
For transform improvement, we design an efficient token mixing block to replace the residual blocks in MLIC$^{++}$. 
Inspired by the MetaFormer architecture~\cite{yu2022metaformer}, {this block employs depth-wise convolution for spatial mixing and lightweight gated interaction for point-wise operations, 
achieving better high-bit-rate performance with lower complexity.}\par
{For entropy modeling, we propose MEMv2, a multi-reference entropy model that captures richer dependencies. 
We introduce hyperprior-guided global correlation prediction for initial slices by exploiting the spatial similarity between hyperprior and latent features. 
A two-stage context modeling strategy is also proposed: it first performs spatial aggregation and then applies adaptive channel-wise reweighting to capture feature importance variations. 
Further enhancements include 2D Rotary Positional Embedding and guided selective compression for improved modeling and coding efficiency.}
\par
{Finally, we employ Stochastic Gumbel Annealing (SGA)~\cite{yang2020improving} to iteratively refine latent representations and side information for each input, 
enabling input-specific rate-distortion optimization without increasing decoder complexity. 
This demonstrates the scalability and adaptability of our framework for future extensions.}
\par 
Our comprehensive approach leads to significant performance gains, as shown in Fig.~\ref{fig:bdbr}. 
MLICv2 achieves BD-rate reductions of 16.54\%, 21.61\%, and 16.05\% on the Kodak, Tecnick, and CLIC Professional Validation datasets, respectively, 
while MLICv2$^+$ further improves these results to 20.46\%, 24.35\%, and 19.14\% over VTM-17.0 Intra.
Our contributions are summarized as follows:
\begin{figure*}[h]
  \centering
  \includegraphics[width=0.95\linewidth]
  {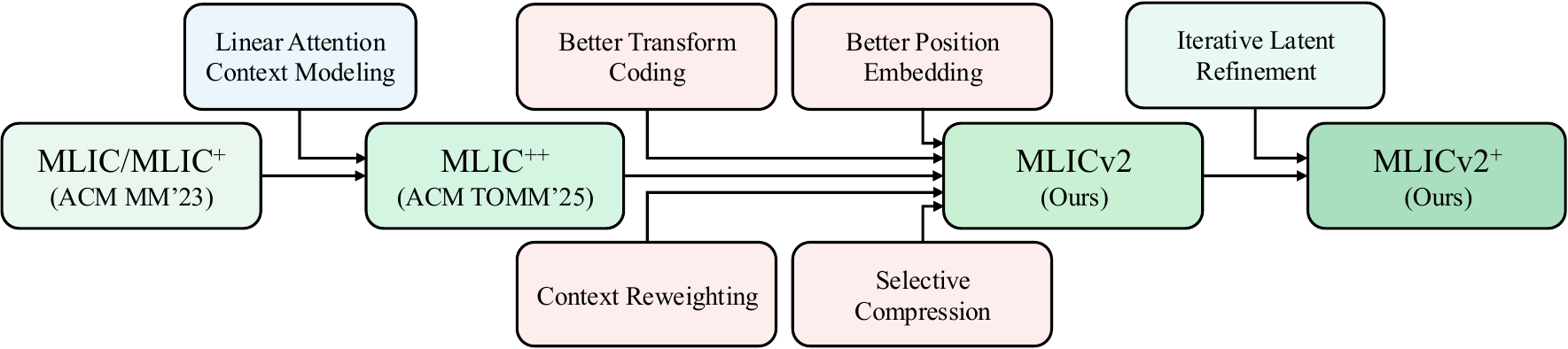}
  \caption{{Model evolution of the MLIC series.}}
  \label{fig:change}
  \end{figure*}
\begin{itemize}
  \item {We introduce a MetaFormer-based transform architecture for LIC through an efficient token mixing block, 
  achieving better high-bit-rate performance and lower complexity than previous residual designs.}
  \item {We enhance entropy modeling with three key innovations:
  (i) hyperprior-guided global correlation prediction for initial slices, 
  (ii) the first two-stage context modeling strategy with spatial aggregation and adaptive channel reweighting, and 
  (iii) advanced positional embedding and selective compression mechanisms.}
  \item {We apply iterative refinement via SGA to demonstrate consistent rate-distortion improvements without increasing decoder complexity.}
  \item {Extensive experiments validate the effectiveness of MLICv2 and MLICv2$^+$, achieving BD-rate reductions of 16.54\%, 21.61\%, 16.05\% and 20.46\%, 24.35\%, 19.14\% 
  on the Kodak, Tecnick, and CLIC Professional Validation datasets, respectively.}
\end{itemize}
\section{Related Works}
\label{sec:related}
\subsection{Learned Image Compression}
Autoencoder-based learned image compression~\cite{balle2016end,theis2017lossy,minnen2018joint,cheng2020learned,he2022elic,jiang2022mlic,jiang2023mlicpp,minnen2020channel,jiang2024llic} 
operates within the transform coding framework, {where an input image 
$\boldsymbol{x} \sim p_{\boldsymbol{x}}$ is first mapped to a latent representation
$\boldsymbol{y} = g_a(\boldsymbol{x}; \phi)$
by an analysis transform $g_a$ with parameters $\phi$, then quantized as
$\hat{\boldsymbol{y}} = \lceil \boldsymbol{y} \rfloor$ 
and finally reconstructed as
$\hat{\boldsymbol{x}} = g_s(\hat{\boldsymbol{y}}; \psi)$
using a synthesis transform $g_s$ with parameters $\psi$. 
A parametric entropy model $p(\cdot; \rho)$ with parameters $\rho$ is employed to estimate the latent distribution 
for efficient compression.
During training, quantization is approximated by adding uniform noise
$\boldsymbol{u} \sim \mathcal{U}(-0.5,0.5)$ or by using the straight-through estimator (STE)~\cite{theis2017lossy}.}
{Following Ball\'e \textit{et al.}~\cite{balle2018variational}, 
the training objective is formulated as:}
\begin{equation}
  {\mathcal{L} = \mathcal{R}(\hat{\boldsymbol{y}}) + \lambda \,\mathcal{D}(\hat{\boldsymbol{x}}, \boldsymbol{x})},
  \end{equation}
{where $\mathcal{R}(\hat{\boldsymbol{y}})$ denotes the estimated bit-rate, $\mathcal{D}$ denotes the reconstruction loss,} and $\lambda$ balances the rate-distortion trade-off.  
\par
Transform architectures have evolved significantly, incorporating Generalized Divisive Normalization (GDN)~\cite{balle2015gdn}, 
{wavelets~\cite{fu2024weconvene,mishra2020wavelet},
attentions~\cite{zhangresidual,chen2021nic,mishra2022deep2}, multi-scale networks~\cite{mishra2022deep}}, deep residual networks~\cite{cheng2020learned}, and transformer-based designs~\cite{zhu2022transformerbased,zou2022the,lu2021tic}.
Mixed CNN-transformer~\cite{liu2023learned} and Mamba~\cite{zeng2025mambaic,gumamba,qin2024mambavc} architectures have also been explored for capturing both local and non-local interactions.
Entropy estimation has progressed from simple univariate Gaussian models~\cite{balle2018variational} to sophisticated approaches including mean-scale Gaussian models~\cite{minnen2018joint}, asymmetric Gaussian distributions~\cite{cui2021asym}, Gaussian mixture models~\cite{cheng2020learned}, and generalized Gaussian models~\cite{zhang2025generalized}, significantly improving compression efficiency.
\subsection{Conditional Entropy Models}
Conditional entropy models leverage context $\hat{\boldsymbol{C}}$ to reduce bit-rate, exploiting the principle that conditional entropy is bounded by unconditional entropy: 
\begin{equation}
  \mathcal{H}(\hat{\boldsymbol{y}}|\hat{\boldsymbol{C}})\leq \mathcal{H}(\hat{\boldsymbol{y}}).
\end{equation}
\par
{Ballé \textit{et al.}~\cite{balle2018variational} introduce hyperprior networks using hyper analysis $h_a$ and hyper synthesis $h_s$ to improve entropy estimation through quantized side information $\hat{\boldsymbol{z}}$ extracted from latent representations.}
\par
{Local context models have evolved from serial methods, such as} PixelCNN~\cite{van2016conditional,minnen2018joint},
to more efficient parallel approaches, including zigzag 
scanning~\cite{li2020efficient} and checkerboard partitioning~\cite{he2022elic},
{which focus on intra-slice dependencies by treating the latent representation as a slice.
Minnen \textit{et al.}~\cite{minnen2020channel} introduced channel-wise autoregressive coding to 
efficiently model dependencies along the channel dimension and achieve faster decoding.}\par
{Recent studies further extend context modeling to capture long-range dependencies~\cite{qian2020learning,guo2021causal,kim2022joint,jiang2022mlic,jiang2023mlicpp}.
For instance, Qian \textit{et al.}~\cite{qian2020learning,qian2022entroformer} introduce bi-directional attention to model global correlations, while Guo \textit{et al.}~\cite{guo2021causal} 
and Kim \textit{et al.}~\cite{kim2022joint} transmit global priors directly to the decoder for enhanced reconstruction.}\par
Recently, Jiang \textit{et al.}~\cite{jiang2022mlic} propose MLIC and MLIC$^+$, 
dividing global contexts into inter-slice and intra-slice components. 
The latent representation is partitioned into multiple slices for autoregressive conditional coding,
employing checkerboard attention for intra-slice contexts and convolutional layers for inter-slice contexts. However, 
the quadratic complexity of vanilla attention~\cite{vas2017attention} limits scalability for high-resolution images, 
which is addressed by MLIC$^{++}$~\cite{jiang2023mlicpp} through linear attention mechanisms~\cite{shen2021efficient}.
\begin{figure*}
\centering
\includegraphics[width=0.95\linewidth]
{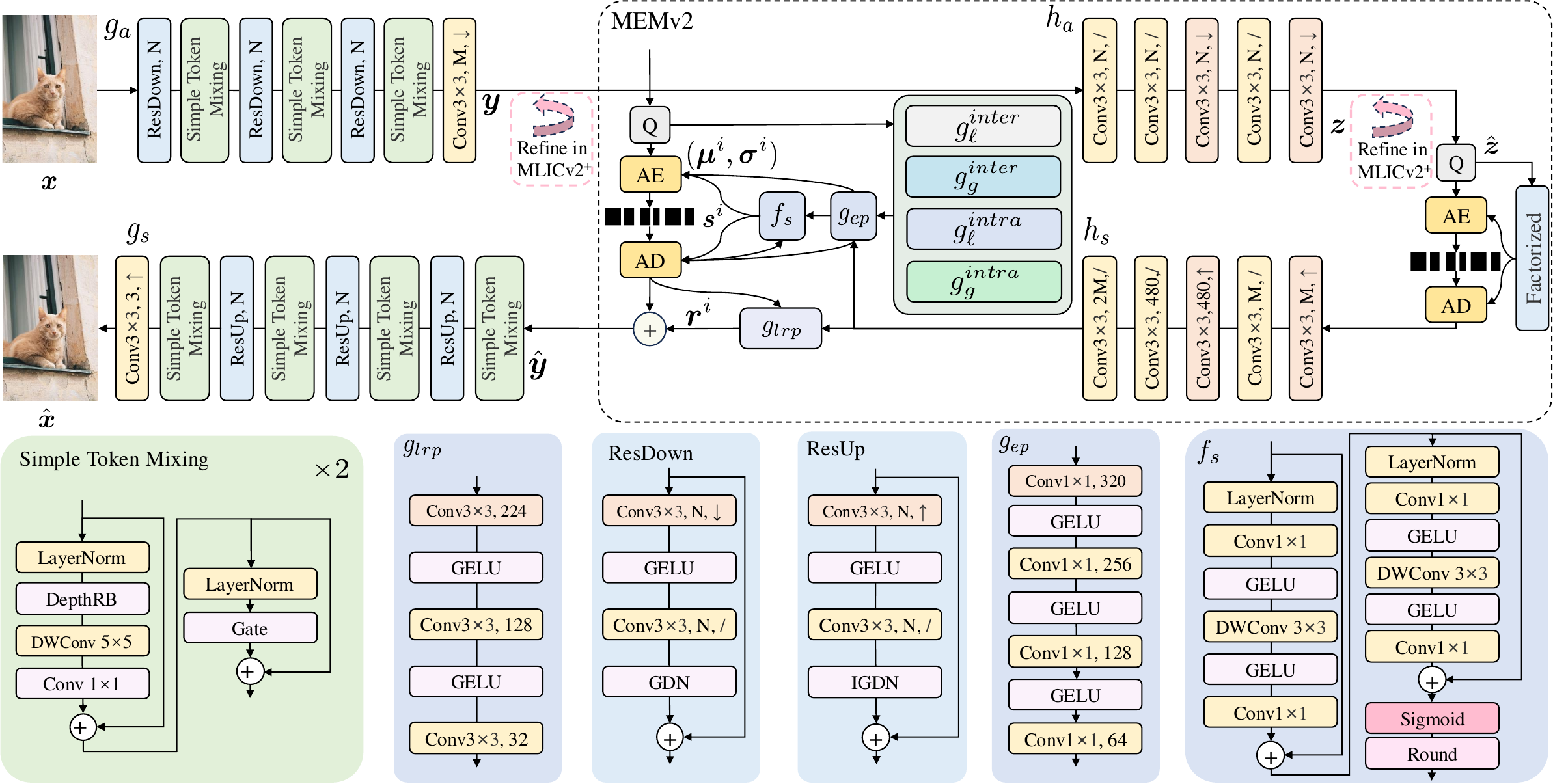}
\caption{The overall architecture of MLICv2/MLICv2$^+$.
$g_a$ is the analysis transform, $g_s$ is the synthesis
transform, $h_a$ is the hyper analysis, and $h_s$ is the hyper synthesis. 
$g_{ep}$ is the entropy parameter module.
$f_s$ is the selective compression predictive module.
$g_{lrp}$ is the latent residual prediction module~\cite{minnen2020channel}. AE and AD are arithmetic encoding and decoding.
$\uparrow$ denotes upsampling and $\downarrow$ denotes downsampling.
${\boldsymbol{x}}$ is the input image and $\hat{\boldsymbol x}$ is the reconstructed image.
$\boldsymbol{y}$ is the latent representation and $\hat{\boldsymbol{y}}$ is
the quantized latent representation. $\hat{\boldsymbol y}^i$ is the $i$-th slice of $\hat{\boldsymbol{y}}$.
$\boldsymbol{\mu}^i, \boldsymbol{\sigma}^i$ are the mean and scale of $\hat{\boldsymbol{y}}^i$. 
$\boldsymbol{s}^i$ is the predicted selective coding map. $\boldsymbol{r}^i$ is the predicted residual.
$\hat{\boldsymbol{z}}$ is the side information. $M, N$ are the channel numbers, which are $320, 192$ in MLICv2/MLICv2$^+$.
``Refine'' is employed in MLICv2$^+$.}
\label{fig:arch}
\end{figure*}
\section{Method}
\label{sec:method}
\subsection{Background: MLIC$^{++}$}
{Our proposed MLICv2 and MLICv2$^{+}$ are built upon the foundation of MLIC$^{++}$~\cite{jiang2023mlicpp}. 
MLIC$^{++}$ adopts a simplified variant of Cheng'20's transform~\cite{cheng2020learned}, 
where attention modules are removed to reduce computational complexity. 
For entropy modeling, MLIC$^{++}$ introduces a linear-complexity multi-reference entropy model (MEM$^{++}$) 
that effectively captures hyperprior, intra-slice, and inter-slice dependencies.}\par
The hyperprior information $\hat{\boldsymbol{H}}$ is derived from side information $\hat{\boldsymbol{z}}$. 
 Following the approach of Minnen \textit{et al.}~\cite{minnen2020channel}, the latent representation $\hat{\boldsymbol{y}}$ is 
 partitioned into $L + 1$ slices: ${\hat{\boldsymbol{y}}^0, \hat{\boldsymbol{y}}^1, \ldots, \hat{\boldsymbol{y}}^L}$. 
 Each slice $\hat{\boldsymbol{y}}^i$ is further divided into anchor $\hat{\boldsymbol{y}}^i_{ac}$ and non-anchor $\hat{\boldsymbol{y}}^i_{na}$ 
 components using checkerboard partitioning~\cite{he2021checkerboard}.\par
 The inter-slice local context module $g_{\ell}^{inter}$ {extracts} inter-slice local context $\hat{\boldsymbol{C}}^{<i}_{\ell}$ 
 from preceding slices $\hat{\boldsymbol{y}}^{<i}$. 
 To model inter-slice global context $\hat{\boldsymbol{C}}^{<i}_{g}$, MLIC$^{++}$ employs linear attention~\cite{shen2021efficient} 
 to achieve an optimal balance between performance and computational complexity. The intra-slice local context $\hat{\boldsymbol{C}}^i_{\ell}$ 
 is extracted from the anchor partition $\hat{\boldsymbol{y}}^i_{ac}$ using overlapped window-based checkerboard attention $g_{\ell}^{intra}$.\par
 A key insight in MLIC$^{++}$ is that feature maps from different slices exhibit similar global similarity patterns, 
 as they can be viewed as thumbnails of the same content. Therefore, the attention maps between anchor 
 and non-anchor parts of the $(i-1)$-th slice are leveraged to predict global similarities for the current slice. 
 The intra-slice global context $\hat{\boldsymbol{C}}_{g}^i$ is extracted from $\hat{\boldsymbol{y}}^i_{ac}$ 
 via the predicted attention map and the intra-slice global context module $g_{g}^{intra}$. This process is formulated as:
 \begin{equation}
  \label{eq:mempp}
  \begin{aligned}
    \hat{\boldsymbol{C}}^{\textless i}_{\ell} &= g_{\ell}^{inter}(\hat{\boldsymbol{y}}^{\textless i}),
    \hat{\boldsymbol{C}}^{\textless i}_{g} = g_{g}^{inter}(\hat{\boldsymbol{y}}^{\textless i}),\\
    \hat{\boldsymbol{C}}^{i}_{\ell} &= g_{\ell}^{intra}(\hat{\boldsymbol{y}}^{i}_{ac}),
    \hat{\boldsymbol{C}}^{i}_{g} = g_{g}^{intra}(\hat{\boldsymbol{y}}^{i-1}_{ac},\hat{\boldsymbol{y}}^{i-1}_{na}, \hat{\boldsymbol{y}}^{i}_{ac}).
  \end{aligned}
  \end{equation}
  The estimated means $\boldsymbol{\mu}^i_{ac}$, $\boldsymbol{\mu}^i_{na}$ and scales $\boldsymbol{\sigma}^i_{ac}$, $\boldsymbol{\sigma}^i_{na}$ 
  are extracted from the contextual information using the entropy parameter module $g_{ep}$. 
  MLIC$^{++}$ incorporates mixed quantization~\cite{minnen2020channel} and the latent residual prediction module $g_{lrp}$~\cite{minnen2020channel}. 
  In mixed quantization, zero-centered {straight-through estimation (STE)~\cite{theis2017lossy}} is applied for distortion computation, 
  while additive uniform noise (AUN)~\cite{balle2016end} is used for entropy estimation during training.
   The module $g_{lrp}$ predicts quantization residuals based on decoded slices and hyperprior information:
   \begin{equation}
    \begin{aligned}
      \hat{\boldsymbol{y}}_{ac}^i &= \textrm{STE}(\boldsymbol{y}^i_{ac} - \boldsymbol{\mu}^i_{ac}) + \boldsymbol{\mu}^i_{ac} + \boldsymbol{r}^i_{ac},\\
      \hat{\boldsymbol{y}}_{na}^i &= \textrm{STE}(\boldsymbol{y}^i_{na} - \boldsymbol{\mu}^i_{na}) + \boldsymbol{\mu}^i_{na} + \boldsymbol{r}^i_{na},\\ 
    \end{aligned}
    \label{eq:ste}
  \end{equation}
  where 
\begin{equation}
  \begin{aligned}
    \boldsymbol{r}^i_{ac} = g_{lrp}(\hat{\boldsymbol{y}}^{\textless i}, \hat{\boldsymbol{H}}),
    \boldsymbol{r}^i_{na} = g_{lrp}(\hat{\boldsymbol{y}}^{\textless i}, \hat{\boldsymbol{y}}^{i}_{ac}, \hat{\boldsymbol{H}}).
  \end{aligned}
\end{equation}
\subsection{Overall Architecture}
  \label{sec:method:overview} 
  Figure~\ref{fig:arch} illustrates the overall architecture of our proposed MLICv2 and MLICv2$^{+}$. 
  {MLICv2 retains the fundamental design of MLIC$^{++}$ but introduces three key enhancements to improve compression efficiency, especially at high bit rates.}
  {First}, residual blocks of transforms are replaced with lightweight token mixing blocks, enhancing representational capacity while reducing computational cost. 
  {Second}, the multi-reference entropy model is upgraded to MEMv2, which integrates improved inter-slice and intra-slice context modules with context reweighting and advanced positional embeddings to better capture latent dependencies. 
{Third}, a selective compression module $f_s$ is introduced to predict and skip zero elements in latent representations, effectively lowering both bit rate and arithmetic coding time. 
{Furthermore}, MLICv2$^{+}$ extends MLICv2 by incorporating iterative refinement of latent features $\boldsymbol{y}$ and side information $\boldsymbol{z}$ according to the rate-distortion characteristics of each input, enabling adaptive and content-specific compression strategies.
  \subsection{Efficient Simple Token Mixing Transform {(STMT)}}
As illustrated in Figure~\ref{fig:rd}, {MLIC$^{++}$ suffers from performance degradation at high bit rates due to the limited capacity of its transform modules 
to retain sufficient information in latent representations~\cite{xu2022multi}. }
The representational capacity is largely determined by the number of channels. 
Ballé \textit{et al.}~\cite{balle2020nonlinear} showed that when channel numbers are small, the performance gap between low- and high-capacity models is negligible at low bit rates but becomes significant as the bit rate increases, with higher-capacity models achieving noticeably better reconstruction quality. 

{Simply enlarging the channel dimension, however, leads to prohibitive computational costs.}
A more effective strategy is to enhance the transform design itself to improve capacity without substantially increasing complexity. 
Recent studies have explored transformer-based transforms~\cite{liu2021swin,liu2023learned,li2024frequencyaware,zou2022the,lu2021tic}, which achieve strong performance gains but often double the computational load compared to conventional residual blocks~\cite{cheng2020learned,jiang2022mlic}. 

Inspired by the MetaFormer framework~\cite{yu2022metaformer}, which separates token mixing and channel interaction, 
we design a lightweight {token mixing block} tailored for learned image compression. 
Unlike MetaFormer's pooling-based token mixing that causes information loss, our 
design adopts a depth-wise residual block (DepthRB)~\cite{jiang2024llic} 
for nonlinear embedding, a depth-wise convolution for spatial mixing, 
and a {gate block~\cite{jiang2024llic}} for efficient channel interactions. 
Layer normalization (LN)~\cite{ba2016layer} is applied for training stability. 
Given an input feature $\boldsymbol{\omega} \in \mathbb{R}^{c \times h \times w}$, the process is formulated as:
\begin{equation}
\begin{aligned}
\boldsymbol{\omega} &= \boldsymbol{\omega} + \mathrm{Conv}_{1\times1}(\mathrm{DWConv}_{5\times5}(\mathrm{DepthRB}(\mathrm{LN}(\boldsymbol{\omega})))),\\
\boldsymbol{\omega} &= \boldsymbol{\omega} + \mathrm{Gate}(\mathrm{LN}(\boldsymbol{\omega})).
\end{aligned}
\end{equation}

Stacking two token mixing blocks yields deeper transforms with enhanced expressiveness. 
Compared to conventional residual blocks with two $3\times3$ convolutions~\cite{cheng2020learned}, 
our design achieves lower complexity and superior rate-distortion performance, particularly at high bit rates. 
Complexity analysis shows that stacking two token mixing blocks reduces MACs by \textbf{7.38\%} at $c=192$ and by \textbf{8.87\%} at $c=320$, compared to a single residual block~\cite{cheng2020learned,jiang2023mlicpp}.
\begin{figure}
  \centering
  \includegraphics[width=0.9\linewidth]
  {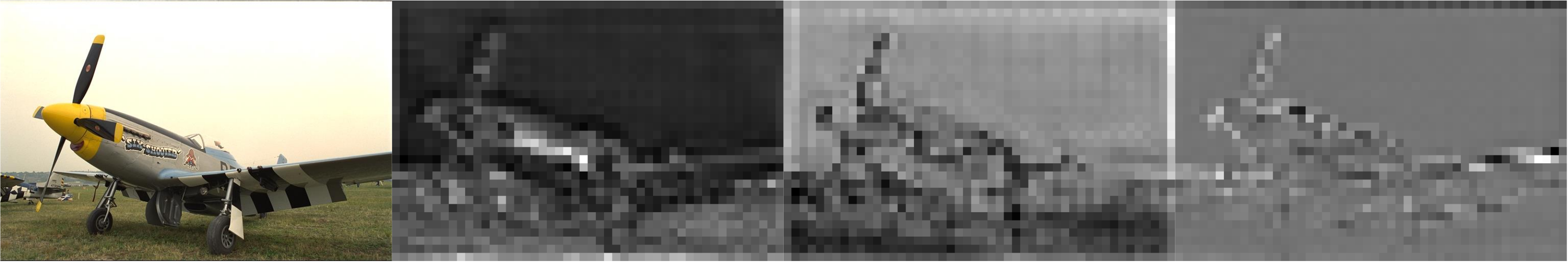}
  \caption{Visualization of the channels of hyperprior $\hat{\boldsymbol{H}}$ ($\lambda=0.013$).
  $\hat{\boldsymbol{H}}$ extracted from the latent representation of Image Kodim20 exhibits similar global similarity to original Kodim20.}
  \label{fig:hyper}
  \end{figure}
\subsection{Enhanced Multi-Reference Entropy Modeling}\label{sec:method:channel}
{The enhanced multi-reference entropy model (MEMv2) is illustrated in Figure~\ref{fig:arch}. 
To overcome the limitations of MLIC$^{++}$, MEMv2 incorporates several key innovations: 
(i) \textbf{hyperprior-guided global correlation prediction (HGCP)} for more accurate dependency modeling, 
(ii) \textbf{context reweighting (CR)} to adaptively emphasize informative channels, 
(iii) \textbf{2D Rotary Positional Embedding (2D RoPE)} for enhanced spatial awareness, and 
(iv) \textbf{guided selective compression (GSC)} to improve entropy coding efficiency. 
Together, these components enable more expressive and adaptive context modeling, particularly at high bit rates.}
\subsubsection{Hyperprior-Guided Global Correlation Prediction {(HGCP)}}
In MLIC$^{++}$ \cite{jiang2023mlicpp}, global similarity between anchor and non-anchor parts of the previous slice is exploited to 
predict current slice similarities. However, for the first slice $\hat{\boldsymbol{y}}^0$, no such reference exists. 
The only available conditional information is the hyperprior $\hat{\boldsymbol{H}}$, obtained by upsampling side information $\hat{\boldsymbol{z}}$ 
extracted from latent representation $\boldsymbol{y}$.
As illustrated in Figure~\ref{fig:hyper}, the hyperprior preserves global similarity patterns similar to the original image. 
We leverage this property by estimating global similarity between $\hat{\boldsymbol{y}}_{ac}^0$ and $\hat{\boldsymbol{y}}_{na}^0$ 
using the similarity between anchor and non-anchor parts of the hyperprior, $\hat{\boldsymbol{H}}_{ac}$ and $\hat{\boldsymbol{H}}_{na}$.
Following MLIC$^{++}$, we employ linear attention~\cite{shen2021efficient} for global context modeling and {employ a Gate block introduced in LLIC~\cite{jiang2024llic} 
for enhanced pointwise interaction}:
\begin{equation}
  \begin{aligned}
    \hat{\boldsymbol{C}}_{g}^0 &= \textrm{Softmax}_2(\hat{\boldsymbol{H}}_{na,q})\left(\textrm{Softmax}_1(\hat{\boldsymbol{H}}_{ac,k})^{\top}\hat{\boldsymbol{y}}^0_{g,v}\right),\\
    \hat{\boldsymbol{C}}_{g}^0 &= \textrm{Gate}(\hat{\boldsymbol{C}}_{g}^0) + \hat{\boldsymbol{C}}_{g}^0,
  \end{aligned}
\end{equation}
where $\hat{\boldsymbol{H}}_{na,q} = {\text{LinearLayer}}(\hat{\boldsymbol{H}}_{na})$, $\hat{\boldsymbol{H}}_{ac,k} =  {\text{LinearLayer}}(\hat{\boldsymbol{H}}_{ac})$, 
and $\hat{\boldsymbol{y}}^0_{g,v} =  {\text{LinearLayer}}(\hat{\boldsymbol{y}}_{ac}^0)$.\par
{While prior methods~\cite{he2022elic,jiang2022mlic,jiang2023mlicpp,jiang2024llic} capture only local dependencies in the first slice, 
our hyperprior-guided mechanism effectively models global dependencies. 
Due to the information compaction property of latent representations~\cite{he2022elic}, 
the first slice typically exhibits higher entropy as it serves as a reference for subsequent slices. 
Incorporating global context at this stage thus significantly reduces bit rate and enhances compression performance.}
\begin{figure}
  \centering
  \includegraphics[width=0.6\linewidth]
  {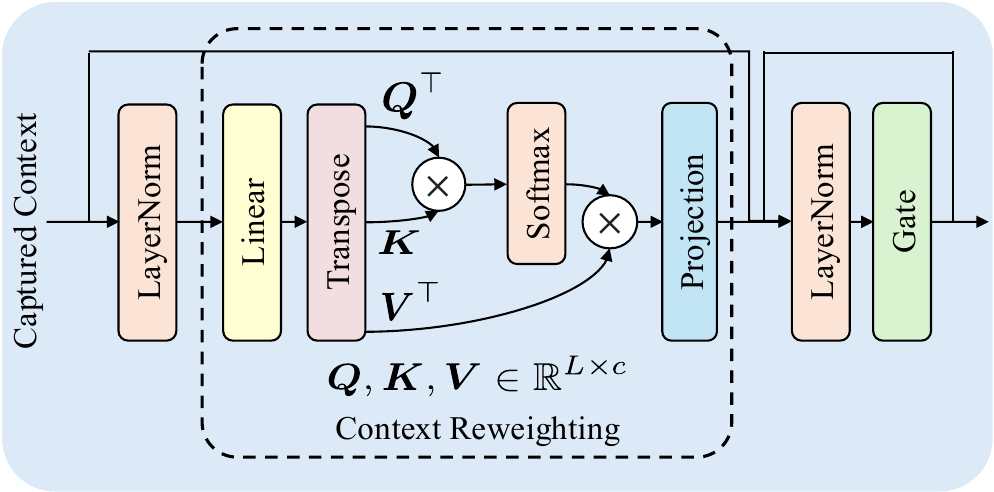}
  \caption{Proposed context reweighting module.}
  \label{fig:memv2}
  \end{figure}
\subsubsection{Context Reweighting {(CR)}}
\label{sec:method:ctx_reweighting}

In MLIC$^{++}$~\cite{jiang2023mlicpp}, the contexts are extracted by spatial attention mechanism. Given the query 
$\boldsymbol{Q}\in \mathbb{R}^{L\times c}$, key $\boldsymbol{K}\in \mathbb{R}^{L\times c}$ and the value 
$\boldsymbol{V}\in \mathbb{R}^{L\times c}$, where $L=hw$ is the sequence length, $h$ is the height, 
$w$ is the width, and $c$ is the channel number, in most spatial attention~\cite{vas2017attention,shen2021efficient}, 
the attention map is computed by $\textrm{Sim}(\boldsymbol{Q}\boldsymbol{K}^{\top}) \in \mathbb{R}^{L\times L}$,
 where $\textrm{Sim}$ 
is the predefined function, such as $\textrm{Softmax}$~\cite{vas2017attention} or two independent $\textrm{Softmax}$ operations~\cite{shen2021efficient}, 
indicating the information is gathered and weighted from 
spatial tokens ($\in \mathbb{R}^{1\times c}$). However, one common limitation is that all channel-wise features are treated equally, 
without considering that different channels may carry varying levels of importance.
We argue that adaptively reweighting channel-wise context features can lead to more accurate entropy modeling. 
By suppressing noisy or less informative channels and emphasizing more relevant ones, 
the model has the potential to capture context information more precisely and improve overall compression performance, 
which has not been considered in existing methods~\cite{jiang2022mlic,jiang2023mlicpp, he2022elic,he2021checkerboard,minnen2020channel,duan2023qarv,fu2023learned,qian2022entroformer,qian2020learning}.
\par
To address this issue, we introduce a two-stage context modeling strategy, where spatial aggregation~\cite{jiang2023mlicpp} is first performed, 
followed by adaptive channel-wise context reweighting to capture the varying importance of different features. The architecture is in Fig.~\ref{fig:memv2}. 
Specifically, in context reweighting, 
the context is transformed to query $\boldsymbol{Q}$, key $\boldsymbol{K}$, and value $\boldsymbol{V}$, 
and the query and values are transposed to compute the channel-wise attention map~\cite{ali2021xcit}. The process is formulated as:
\begin{equation}
\boldsymbol{O} = \textrm{Softmax}(\boldsymbol{Q}^{\top}\boldsymbol{K})\boldsymbol{V}^{\top}.
\end{equation}
{The attention map $\boldsymbol{M}=\textrm{Softmax}(\boldsymbol{Q}^{\top}\boldsymbol{K})$ is row-wise normalized via the Softmax function:
$
    \boldsymbol{M}_{i,j} = \frac{\exp(\boldsymbol{q}_i^\top \boldsymbol{k}_j)}{\sum_{j^{\ast}=1}^C\exp(\boldsymbol{q}_i^\top \boldsymbol{k}_{j^\ast})},
$ 
  where $\boldsymbol{q}_i$ and $\boldsymbol{k}_j$ are the $i$-th and $j$-th channel 
  vectors of $\boldsymbol{Q}$ and $\boldsymbol{K}$, respectively.
  The summation is over spatial indices while channel indices remain open.} 
The resolution of the attention map $\textrm{Softmax}(\boldsymbol{Q}^{\top}\boldsymbol{K})$ is $c\times c$, 
indicating the importance of each channel ($\in \mathbb{R}^{L\times 1}$). 
The complexity of Eq. (8) is $\mathcal{O}(c^2hw)$. $\boldsymbol{O}$ is then fed into a Gate block~\cite{jiang2024llic}
 for efficient point-wise interactions. This attention-based reweighting helps the model focus on the most relevant parts of the context. 
 To the best of our knowledge, this is the first attempt to explicitly model channel-wise importance in context features for LIC.
  Our approach brings consistent performance improvements with minimal additional cost.
\subsubsection{2D Rotary Position Embedding {(2D RoPE)}}
MLIC$^{++}$~\cite{jiang2023mlicpp} applies relative position embedding (RPE) from Swin-Transformer~\cite{liu2021swin}
 directly to attention maps as additional bias. However, this direct addition may limit interaction with attention weights. 
 Recently, Rotary Positional Embedding (RoPE)~\cite{su2024roformer}, a multiplication-based approach, 
 has been widely adopted in language models~\cite{touvron2023llama,reid2024gemini}.
RoPE offers several advantages: compatibility with linear attention (whereas RPE requires adding biases that 
conflict with implicit attention computation), better extrapolation capabilities, and more concise implementation.
For 2D image content, we extend the original 1D RoPE to handle horizontal {(width)}, vertical {(height)}, and diagonal relationships. 
Given query $\boldsymbol{q}_{m_x,m_y}$ and key $\boldsymbol{k}_{n_x,n_y}$ at 2D positions $(m_x, m_y)$ and $(n_x, n_y)$, 
we rotate queries and keys twice with different angles $\theta_x$ and $\theta_y$ for horizontal and vertical directions.
When $c=2$, the rotation matrix for the query becomes:
\begin{equation}
  \begin{aligned}
    \underbrace{\begin{bmatrix}\cos m_x\theta_x & -\sin m_x\theta_x\\ \sin m_x\theta_x & \cos m_x\theta_x\end{bmatrix}}_{\textrm{horizontal}}\underbrace{\begin{bmatrix}\cos m_y\theta_y & -\sin m_y\theta_y\\ \sin m_y\theta_y & \cos m_y\theta_y\end{bmatrix}}_{\textrm{vertical}}
  =\begin{bmatrix}\cos (m_x\theta_x + m_y\theta_y) & -\sin (m_x\theta_x + m_y\theta_y)\\ \sin (m_x\theta_x + m_y\theta_y) & \cos (m_x\theta_x + m_y\theta_y)\end{bmatrix}.
  \end{aligned}
\end{equation}
The attention score computation yields:
\begin{equation}
  \begin{aligned}
    &\boldsymbol{q}_{m_x,m_y}^{\prime} = \boldsymbol{q}_{m_x,m_y}\begin{bmatrix}\cos (m_x\theta_x + m_y\theta_y) & -\sin (m_x\theta_x + m_y\theta_y)\\ \sin (m_x\theta_x + m_y\theta_y) & \cos (m_x\theta_x + m_y\theta_y)\end{bmatrix},\\
    &\boldsymbol{k}_{n_x,n_y}^{\prime} = \boldsymbol{k}_{n_x,n_y}\begin{bmatrix}\cos (n_x\theta_x + n_y\theta_y) & -\sin (n_x\theta_x + n_y\theta_y)\\ \sin (n_x\theta_x + n_y\theta_y) & \cos (n_x\theta_x + n_y\theta_y)\end{bmatrix},\\
    &\boldsymbol{q}_{m_x,m_y}^{\prime}(\boldsymbol{k}_{n_x,n_y}^{\prime})^{\top} = \boldsymbol{q}_m\begin{bmatrix}\cos\theta_{(m_x-n_x, m_y - n_y)}  & -\sin \theta_{(m_x-n_x, m_y - n_y)}\\ \sin \theta_{(m_x-n_x, m_y - n_y)} & \cos \theta_{(m_x-n_x, m_y - n_y)}\end{bmatrix}\boldsymbol{k}_n^{\top},
  \end{aligned}
  \label{eq:rope2d}
\end{equation}
where $\theta_{(m_x-n_x, m_y-n_y)} = (m_x-n_x)\theta_x + (m_y-n_y)\theta_y$ encodes relative positions.
Due to the ambiguity in defining relative importance of horizontal versus vertical positional dependencies, we make angles $\theta_x$ and $\theta_y$ learnable, initializing them to 10000 following original RoPE~\cite{su2024roformer}.
{For $c \textgreater 2$, the queries and keys are divided into $\frac{c}{2}$ groups,
each following Equation~(\ref{eq:rope2d})}.
\subsubsection{Guided Selective Compression {(GSC)}}
Zero-centered quantization in Equation~\ref{eq:ste} produces many zero elements in 
$\lceil\boldsymbol{\Lambda}\rfloor = \lceil\boldsymbol{y} - \boldsymbol{\mu}\rfloor$, 
as illustrated in Figure~\ref{fig:ele}. While these zeros consume minimal bit rate, 
they account for significant arithmetic encoding/decoding time.

Recent selective compression methods~\cite{lee2022selective,alexandre2023hierarchical} have employed neural networks to 
predict selective compression maps, but they are primarily applied to simpler entropy models. In addition, 
scale-based prediction has been introduced in recent works~\cite{shi2022alphavc,zhang2024practical}. 
However, when extended to advanced entropy models~\cite{jiang2022mlic,jiang2023mlicpp}, 
\textit{the training phase becomes highly unstable, 
with the overall loss value becoming ``nan" after several steps, 
regardless of whether the parameters of the compression model are frozen or not.}
{The observed training instability is primarily caused by the dependency introduced by the context modeling, 
rather than by the selective compression mechanism itself. Specifically, skipping non-zero elements in the current slice can 
significantly 
affect the entropy modeling of subsequent slices due to the inter-latent dependencies established by the context model.
Considering the substantial performance gains enabled by context modeling, 
our goal is to skip the encoding of zero elements as much as 
possible—on top of the existing context modeling framework~\cite{minnen2020channel,he2022elic,jiang2023mlicpp,jiang2022mlic}—in order 
to reduce the time spent on arithmetic encoding and decoding.}\par
{
To address above issues, the guided selective compression module $f_s$ is proposed for post-training.}
This module is employed on the converged MLICv2, with the weights of MLICv2 frozen for stable training. 
The architecture of the guided selective compression $f_s$ module is presented in Fig.~\ref{fig:arch}.
Specifically, in $f_s$, the scale-based selective prediction is employed as the initial predicted 
selective compression map. For $j$-th element ${s}_{\boldsymbol{\sigma},j}$ in $\boldsymbol{s}_{\boldsymbol{\sigma}}$, 
the scale-based selective compression value is {served as the initialization}:
\begin{equation}
    \label{eq5}
    {s}_{\boldsymbol{\sigma},j} = \left\{
    \begin{aligned}
    &0, \quad {\sigma}_j \textless {\xi}, \\
    &1, \quad {\sigma}_j \geq {\xi},
    \end{aligned}
    \right.
\end{equation}
where $\xi$ is a predefined threshold, set to $0.3$.
For $\boldsymbol{y}^i_{ac}$, the decoded elements of previous slices serve as priors.
For $\boldsymbol{y}^i_{na}$, the decoded elements of previous slices and $\boldsymbol{y}^i_{ac}$
serve as priors. The hyperprior $\hat{\boldsymbol{H}}$
also serves as a prior. {The scale-based initialization is refined by $f_s$ for final skip map:}
\begin{equation}
  \begin{aligned}
    \boldsymbol{\varepsilon}^i_{ac} &= {\textrm{Sigmoid}}\left(f_s\left(\boldsymbol{s}_{\boldsymbol{\sigma},ac}^i, \lceil\boldsymbol{\Lambda}^{\textless i}\rfloor\odot \boldsymbol{s}^{\textless i}, \hat{\boldsymbol{H}}\right)\right),
    \boldsymbol{s}^i_{ac} = \textrm{STE}\left(\boldsymbol{\varepsilon}^i_{ac}\right),\\
    \boldsymbol{\varepsilon}^i_{na} &= {\textrm{Sigmoid}}\left(f_s\left(\boldsymbol{s}_{\boldsymbol{\sigma},na}^i, \lceil\boldsymbol{\Lambda}^{\textless i}\rfloor \odot \boldsymbol{s}^{\textless i},
    \lceil\boldsymbol{\Lambda}^i_{ac}\rfloor\odot \boldsymbol{s}^{i}_{ac}, \hat{\boldsymbol{H}} \right)\right),
    \boldsymbol{s}^i_{na} = \textrm{STE}\left(\boldsymbol{\varepsilon}^i_{na}\right).
  \end{aligned}
\end{equation}
{During training, $\lceil\boldsymbol{\Lambda}\rfloor$ is known, therefore 
cross-entropy loss is employed to optimize $f_s$,
treating selective prediction as a \textit{binary classification} task.
Additionally, elements with larger values have a higher impact on distortion, 
so the cross-entropy loss is weighted by {$|\lceil\boldsymbol{\Lambda}\rfloor|+1$}.
The overall loss is:
\begin{figure}
  \centering
  \includegraphics[width=0.7\linewidth]
  {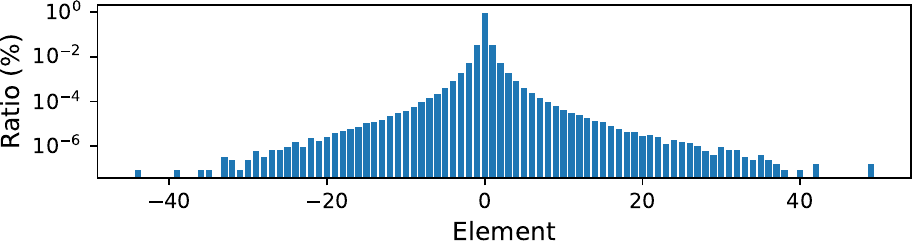}
  \caption{Ratios of different elements in $\lceil\boldsymbol{\Lambda}\rfloor=\lceil\boldsymbol{y} - \boldsymbol{\mu}\rfloor$ when the model is optimized for MSE and the $\lambda=0.013$.
  There are many zero elements in $\lceil\boldsymbol{\Lambda}\rfloor$.}
  \label{fig:ele}
  \end{figure}
  \begin{equation}
    \begin{aligned}
      &{\mathcal{L}_s = \sum_{i=0}^L \Bigl((|\lceil\boldsymbol{\Lambda}^i_{ac}\rfloor| + 1)\odot \mathcal{L}^{i}_{ce,ac} +
      (|\lceil\boldsymbol{\Lambda}^i_{na}\rfloor| + 1)\odot \mathcal{L}^{i}_{ce,na}\Bigr)},\\
      &{\mathcal{L}_{ce,ac}^i = \tfrac{1}{2}\!\left[-\mathbf{1}\!\big(|\lceil\boldsymbol{\Lambda}^i_{ac}\rfloor| > 0\big) 
      \odot \log\boldsymbol{\varepsilon}^i_{ac}
      -\big(1-\mathbf{1}\!\big(|\lceil\boldsymbol{\Lambda}^i_{ac}\rfloor| > 0\big)\big)\odot\log(1-\boldsymbol{\varepsilon}^i_{ac})\right]},\\
      &{\mathcal{L}_{ce,na}^i = \tfrac{1}{2}\!\left[-\mathbf{1}\!\big(|\lceil\boldsymbol{\Lambda}^i_{na}\rfloor| > 0\big) \odot \log\boldsymbol{\varepsilon}^i_{na}
      -\big(1-\mathbf{1}\!\big(|\lceil\boldsymbol{\Lambda}^i_{na}\rfloor| > 0\big)\big)\odot\log(1-\boldsymbol{\varepsilon}^i_{na})\right]}.
    \end{aligned}
    \label{eq:skip_refined}
    \end{equation}
    {Since both the encoder and decoder compute identical selective compression maps using $f_s$ with the same inputs, the selective coding decisions are perfectly synchronized without requiring any additional information transmission or bitstream overhead.}
    To the best of our knowledge, we are the
    {\textit{first}} to highlight the importance of quantization residuals and 
    incorporate a weighted loss accordingly.}
\subsection{Iterative Latent Refinement {(ILR)}}
To validate our MLICv2's potential and demonstrate achievable performance improvements, we employs
\textit{encoding-time scaling} to find optimal quantized latent representation $\hat{\boldsymbol{y}}$ and 
side information $\hat{\boldsymbol{z}}$ for each input. 
Direct encoder optimization requires 
storing model weights, gradients and intermediate features, leading to prohibitive memory consumption for high-resolution images.
We adopt Stochastic Gumbel Annealing (SGA)~\cite{yang2020improving} to refine $\boldsymbol{y}$ and $\boldsymbol{z}$ efficiently. 
In SGA, outputs of $g_a$ and $h_a$ serve as initialization. For anchor part $\boldsymbol{y}^i_{ac}$, $\boldsymbol{\Lambda}^i_{ac} = \boldsymbol{y}^i_{ac} - \boldsymbol{\mu}^i_{ac}$ is stochastically rounded up or down.
The stochastic rounding direction is sampled from a relaxed one-hot categorical distribution:
\begin{equation}
  \label{eq6}
  \mathit{q}_{\tau}(\boldsymbol{o}^i_{ac}|\boldsymbol{\Lambda}^i_{ac})\propto\left\{
  \begin{aligned}
  \exp\left\{ \frac{-\textrm{atanh}(\boldsymbol{\Lambda}^i_{ac} - \lfloor \boldsymbol{\Lambda}^i_{ac} \rfloor)}{\tau} \right\}, \boldsymbol{o}^i_{ac}=\{0,1\}, \\
  \exp\left\{ \frac{-\textrm{atanh}(\lceil \boldsymbol{\Lambda}^i_{ac} \rceil - \boldsymbol{\Lambda}^i_{ac})}{\tau} \right\}, \boldsymbol{o}^i_{ac}=\{1,0\},
  \end{aligned}
  \right.
  \end{equation}
where temperature $\tau = \min(0.5, e^{0.001j})$ decreases with refinement step $j$, gradually making the latent representation aware of quantization.
Notably, selective compression is not incorporated during SGA. Instead, the selective compression map is predicted by $f_s$ based on the overfitted latent representations obtained after SGA.
we primarily employ it as a validation tool to demonstrate our model's potential for instance-level adaptability 
and performance improvements.
Rate control mechanisms~\cite{zhang2024practical} could be incorporated in future work for applications requiring strict bit-rate constraints.
\begin{figure*}[t]
  \centering
  \includegraphics[width=\linewidth]
    {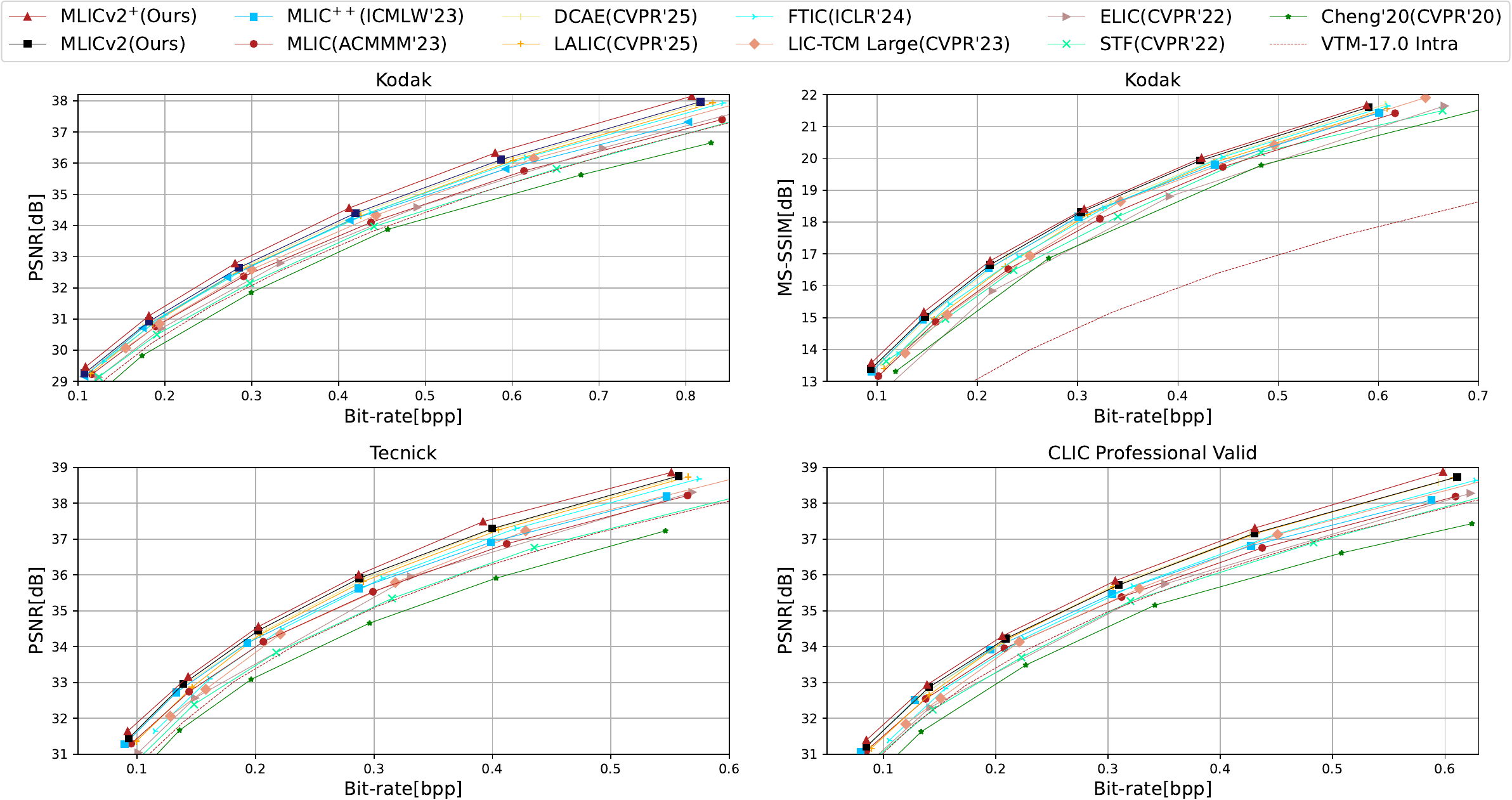}
  \caption{{PSNR-Bit-rate curve and MS-SSIM-Bit-rate curve.
  MS-SSIM is converted to dB for better visual clarity. Please zoom in for better view.}}
  \label{fig:rd}
\end{figure*}
\begin{table*} 
  \centering
  \footnotesize
  \setlength{\tabcolsep}{3.5mm}{
  \caption{{BD-Rate $(\%)$ comparison for PSNR (dB) and MS-SSIM. The anchor is VTM-17.0 Intra. ``$-$'' means the result is not available.}}
  \label{tab:rd} 
  \renewcommand\arraystretch{1.19}
  \begin{tabular}{@{}cccccccccccccccc@{}}
  \toprule
  \multicolumn{1}{c}{\multirow{3}{*}{Methods}} & \multicolumn{1}{c|}{\multirow{3}{*}{{Venue}}}                            & \multicolumn{4}{c}{{BD-Rate (\%) w.r.t. VTM 17.0 Intra}}  \\
  \multicolumn{1}{c}{}      & \multicolumn{1}{c|}{}                                                & \multicolumn{2}{c}{Kodak~\cite{kodak}}    & \multicolumn{2}{c}{Tecnick~\cite{tecnick2014TESTIMAGES}}  & \multicolumn{2}{c}{CLIC Pro Valid~\cite{CLIC2020}}   \\
  \multicolumn{1}{c}{}     & \multicolumn{1}{c|}{}                                                   & \multicolumn{1}{c}{PSNR} & \multicolumn{1}{c}{MS-SSIM}& \multicolumn{1}{c}{PSNR} & \multicolumn{1}{c}{{MS-SSIM}} & \multicolumn{1}{c}{PSNR} & \multicolumn{1}{c}{{MS-SSIM}}  \\ \midrule
  \multicolumn{1}{c|}{VTM-17.0 Intra~\cite{bross2021vvc}}                                        & $0.00$       & $0.00$ & $0.00$       & $0.00$ & $0.00$       & $0.00$    \\\midrule
  \multicolumn{1}{c}{Cheng'20~\cite{cheng2020learned}}     & \multicolumn{1}{c|}{CVPR'20}                                  & 5.58       & -44.21 & 7.57   & {-39.61}    & {11.71}    & {-41.29}    \\
  \rowcolor[HTML]{E6E6E6}\multicolumn{1}{c}{ChARM~\cite{minnen2020channel}}    & \multicolumn{1}{c|}{ICIP'20}                                       & 3.23       & - & -0.88  & -     & - & -    \\
  \multicolumn{1}{c}{Qian'21~\cite{qian2020learning}}      & \multicolumn{1}{c|}{ICLR'21}                                     & 10.05       & -39.53 & 7.52   & -    & 0.02   & -  \\
  \rowcolor[HTML]{E6E6E6}\multicolumn{1}{c}{Xie'21~\cite{xie2021enhanced}}        & \multicolumn{1}{c|}{ACMMM'21}                                   & 1.55       & -43.39 &  -0.80 & -   & 3.21   & -   \\
  \multicolumn{1}{c}{NLAIC'21~\cite{chen2021nic}}        & \multicolumn{1}{c|}{TIP'21}                                   & 11.80       & -40.27 &  7.11 & -   & -  & -    \\
  \rowcolor[HTML]{E6E6E6}\multicolumn{1}{c}{Gao'21~\cite{gao2021neural}}        & \multicolumn{1}{c|}{ICCV'21}                                   & -8.76       & -48.60 &  -   & - & -6.49    & -  \\
  \multicolumn{1}{c}{Guo'22~\cite{guo2021causal}}   & \multicolumn{1}{c|}{TCSVT'22}                                       & -4.45       & -45.23 & -  & -     & -    & -  \\
  \rowcolor[HTML]{E6E6E6}\multicolumn{1}{c}{LBHIC~\cite{wu22block}}   & \multicolumn{1}{c|}{TCSVT'22}                                       & -4.56       & -50.54 & -   & -    & -  & -    \\
  \multicolumn{1}{c}{Entroformer~\cite{qian2022entroformer}}   & \multicolumn{1}{c|}{ICLR'22}                                       & 4.73       & -42.64 & 2.31   & -    & -1.04  & -   \\
  \rowcolor[HTML]{E6E6E6}\multicolumn{1}{c}{SwinT-ChARM~\cite{zhu2022transformerbased}}    & \multicolumn{1}{c|}{ICLR'22}              & -1.73      & -   & -& -& -& - \\
  \multicolumn{1}{c}{NeuralSyntax~\cite{wang2022neural}}   & \multicolumn{1}{c|}{CVPR'22}       & 8.97      & -39.56  & -  & -    & 5.64  &{-38.92}  \\          
  \rowcolor[HTML]{E6E6E6}\multicolumn{1}{c}{STF~\cite{zou2022the}}   & \multicolumn{1}{c|}{CVPR'22}       & -2.48      & -47.72  & -2.75  & -    & 0.42 & -  \\          
  \multicolumn{1}{c}{WACNN~\cite{zou2022the}}   & \multicolumn{1}{c|}{CVPR'22}       & -2.95      & -47.71  & -5.09 & -     & 0.04  & -   \\          
  \rowcolor[HTML]{E6E6E6}\multicolumn{1}{c}{ELIC~\cite{he2022elic}}         & \multicolumn{1}{c|}{CVPR'22}                     & -5.95      & -44.60   & -9.14  & -    & -3.45 & -  \\
  \multicolumn{1}{c}{Contextformer~\cite{koyuncu2022contextformer}}  & \multicolumn{1}{c|}{ECCV'22}      & -5.77      & -46.12 & -9.05 & {-42.29}    & - & -  \\
  \rowcolor[HTML]{E6E6E6}\multicolumn{1}{c}{GLLMM~\cite{fu2023learned}} & \multicolumn{1}{c|}{TIP'23}       & -1.43      & -47.41 & -5.58   & -   & -  & -   \\
  \multicolumn{1}{c}{NVTC~\cite{feng2023nvtc}} & \multicolumn{1}{c|}{CVPR'23}       & -1.04      & -& - & -      & -3.61 & -    \\
  \rowcolor[HTML]{E6E6E6}\multicolumn{1}{c}{LIC-TCM Large~\cite{liu2023learned}}   & \multicolumn{1}{c|}{CVPR'23}      & -10.14      & -48.94 & -11.47  & -    & -8.04& -    \\
  \multicolumn{1}{c}{QARV~\cite{duan2023qarv}}     & \multicolumn{1}{c|}{TPAMI'24}                                  & 0.31       & - & -3.03       & {-}        \\
  \rowcolor[HTML]{E6E6E6}\multicolumn{1}{c}{FTIC~\cite{li2024frequencyaware}}   & \multicolumn{1}{c|}{ICLR'24}      & -12.99      & -51.13 & -14.88   & {-48.42}   & -9.53  & {-45.98}  \\
  \multicolumn{1}{c}{LLIC-TCM~\cite{jiang2024llic}}   & \multicolumn{1}{c|}{TMM'24}      & -10.94      & -49.73 & -14.99   & -   & -10.41 & -   \\
  \rowcolor[HTML]{E6E6E6}\multicolumn{1}{c}{WeConvene~\cite{fu2024weconvene}}   & \multicolumn{1}{c|}{ECCV'24}      & -6.71      & -48.63 & -8.42    & -  & -5.74  & -  \\
  \multicolumn{1}{c}{CCA~\cite{han2024causal}}   & \multicolumn{1}{c|}{NeurIPS'24}      & -12.04      & - & -14.55   &-   & -10.75    \\
  \rowcolor[HTML]{E6E6E6}\multicolumn{1}{c}{MambaIC~\cite{zeng2025mambaic}}   & \multicolumn{1}{c|}{CVPR'25}      & -13.03      & - & -18.41  & -    & -15.58   & - \\
  \multicolumn{1}{c}{LALIC~\cite{feng2025linear}}   & \multicolumn{1}{c|}{CVPR'25}      & -13.72      & -50.86 & -18.19   & -   & -14.34 & -   \\
  \rowcolor[HTML]{E6E6E6}\multicolumn{1}{c}{DCAE~\cite{lu2025learned}}   & \multicolumn{1}{c|}{CVPR'25}      & -15.43      & -53.20 & -20.51   & {-53.45}   & -15.78  & {-50.93}  \\
  \midrule
  \textit{\textbf{Baseline Models}} \\
  \multicolumn{1}{c}{MLIC~\cite{jiang2022mlic}}     & \multicolumn{1}{c|}{ACMMM'23} & {-8.05}      & -49.13 & -12.73  & {-47.26}      & -8.79   & {-45.79}   \\
  \multicolumn{1}{c}{MLIC$^{+}$~\cite{jiang2022mlic}  }     & \multicolumn{1}{c|}{ACMMM'23}   &{-11.39}      & -52.75 & -16.38  & {-53.54}   & -12.56   & {-48.75} \\
  \multicolumn{1}{c}{MLIC$^{++}$~\cite{jiang2023mlicpp}}  & \multicolumn{1}{c|}{ICMLW'23}     & {-13.39}      & {-53.63} & {-17.59} & {-53.83}    & {-13.08}   & {-50.78}  \\\midrule
  \rowcolor[HTML]{E6E6E6}\multicolumn{1}{c}{MLICv2 }     & \multicolumn{1}{c|}{Ours}   & \textbf{-16.54}      & \textbf{-54.56} & \textbf{-21.61}  & {\textbf{-55.26}}    & \textbf{-16.05}   & {\textbf{-51.86}}   \\
  \rowcolor[HTML]{E6E6E6}\multicolumn{1}{c}{MLICv2$^{+}$ }      & \multicolumn{1}{c|}{Ours}   & \textbf{-20.46}      & \textbf{-55.86} & \textbf{-24.35}  & {\textbf{-56.42}}   & \textbf{-19.14}  & {\textbf{-52.75}}  \\\bottomrule  
\end{tabular}}
\end{table*}
\section{Experiments}
\label{sec:exp}
\subsection{Implementation Details}
\label{sec:exp:setup}
\subsubsection{Training Dataset}
Our training dataset contains 101,104 images with resolutions exceeding $640\times 640$ pixels.
These images are carefully selected from various established 
image datasets~\cite{deng2009imagenet,lin2014microsoft,Agustsson2017NTIRE2C,lim2017enhanced,yang2023hq}. 
To address the existing compression artifacts in JPEG images, 
we follow the approach of Ballé \textit{et al.}~\cite{balle2018variational} by further down-sampling the JPEG images using a randomized factor. 
Additionally, images with bits per pixel (bpp) lower than 3 are excluded to ensure data quality.
\subsubsection{Training Strategy}\label{sec:exp:imp:strategy}
MLICv2 is implemented using PyTorch 2.2.2 and CompressAI 1.2.6. 
Following CompressAI, we configure $\lambda \in \{18, 35, 67, 130, 250, 483\} \times 10^{-4}$ for MSE optimization 
and $\lambda \in \{2.4, 4.58, 8.73, 16.64, 31.73, 60.5\}$ for Multi-Scale Structural Similarity (MS-SSIM)~\cite{wang2003multiscale}.
Training is conducted on eight Tesla A100-80G GPUs with a batch size of 16, employing the Adam optimizer $\beta_1=0.9, \beta_2=0.999$. 
Our training consists of two stages:
\textit{Stage 1:} We train the model with only hyperprior for 2 million (M) steps using a learning rate of $10^{-4}$ and $256\times 256$ patches, 
providing superior initialization for analysis and synthesis transforms.
\textit{Stage 2:} Loading the pre-trained weights, we train the complete model for 2M steps. 
The learning rate follows a scheduled decay: starting at $10^{-4}$, then decreasing to $3\times 10^{-5}$, $10^{-5}$, $3\times 10^{-6}$ and $10^{-6}$,
at 1.5M, 1.8M, 1.9M, and 1.95M steps, respectively. We employ $256\times 256$ patches for the initial 1.2M steps, 
then transition to $512\times 512$ patches to leverage the effectiveness of global context modules.
For the guided selective compression module optimization, we conduct 20K steps with batch size 16 and learning rate  $10^{-3}$.\par
To validate our model's potential for further improvement, we employ SGA in MLICv2$^+$,
refining latent representations and side information for 3000 steps with learning rate $10^{-3}$. 
This approach allows us to further enhance the performance under our architectural framework.
\subsection{Benchmarks and Metrics}
We evaluate rate-distortion performance on three datasets extensively used in learned compression research~\cite{balle2016end,theis2017lossy,balle2018variational,minnen2018joint,minnen2020channel,
cheng2020learned,xie2021enhanced,he2021checkerboard,he2022elic,zou2022the,
zhu2022transformerbased,koyuncu2022contextformer,duan2023qarv,wang2022neural,guo2021causal,wu22block,
chen2021nic,liu2023learned,jiang2022mlic,jiang2023mlicpp,jiang2024llic}:
\begin{itemize}
\item Kodak~\cite{kodak}: Contains 24 uncompressed images with $768\times 512$ pixels.
\item Tecnick~\cite{tecnick2014TESTIMAGES}: Comprises 100 uncompressed images with $1200\times 1200$ pixels.
\item CLIC Pro Valid~\cite{CLIC2020}: The validation set from the 3rd Challenge on Learned Image Compression, 
consisting of 41 high-resolution images with $2048\times1440$ pixels.
\end{itemize}
Performance ranking employs the Bjøntegaard delta rate (BD-Rate) metric~\cite{bjontegaard2001calculation} for comprehensive evaluation.
\par
\begin{figure*}
  \centering
  \includegraphics[width=0.9\linewidth]
  {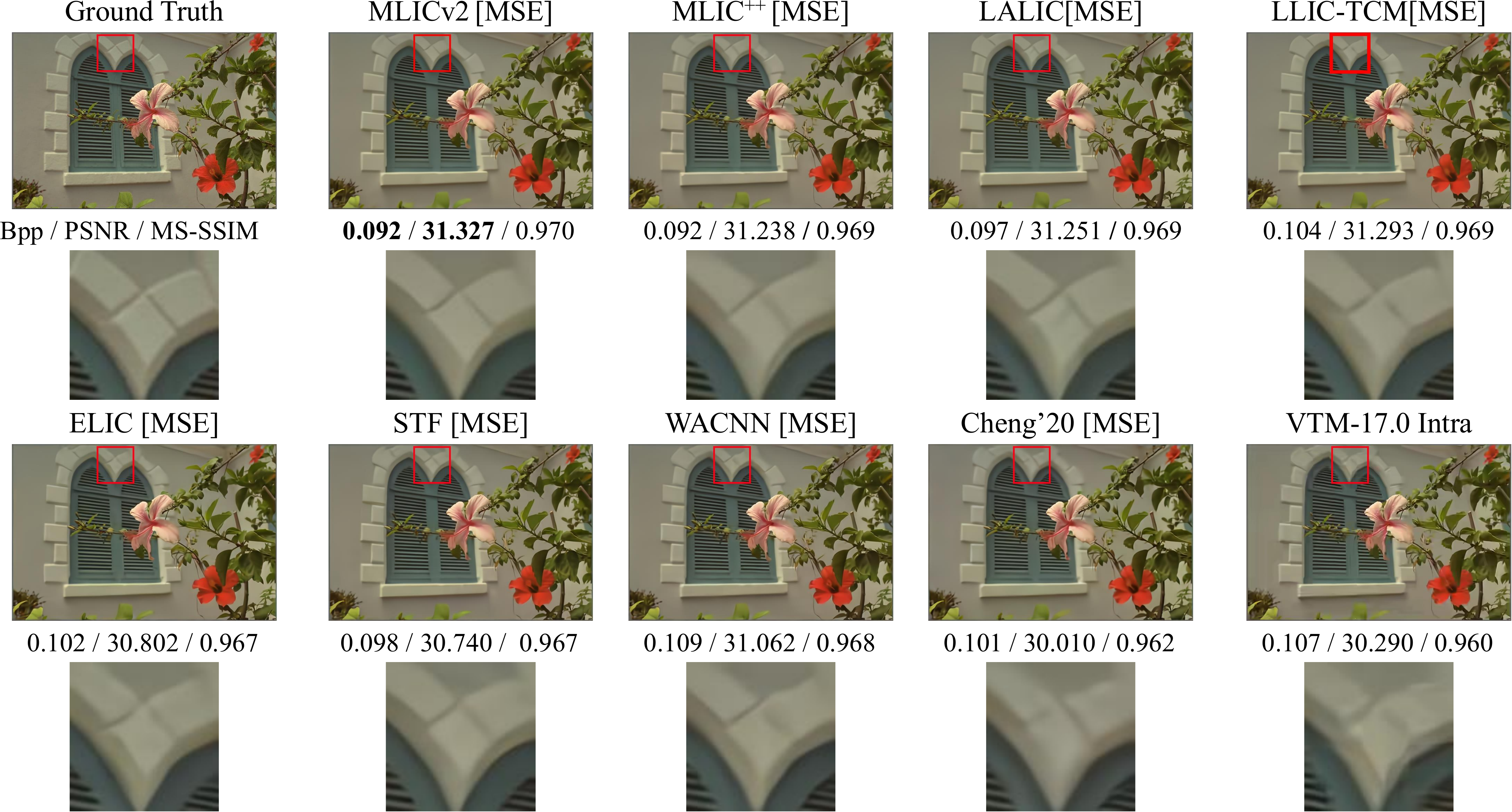}
  \caption{{Reconstructions of our proposed MLICv2, recent learned image compression models~\cite{jiang2023mlicpp,he2022elic,zou2022the,cheng2020learned,feng2025linear,jiang2024llic} and VTM-17.0 Intra~\cite{bross2021vvc}.
  “[MSE]” denotes the model is optimized for MSE. Please zoom in for better view.}}
  \label{fig:vis}
  \end{figure*}
\begin{figure*}
      \centering
    \includegraphics[scale=0.5]{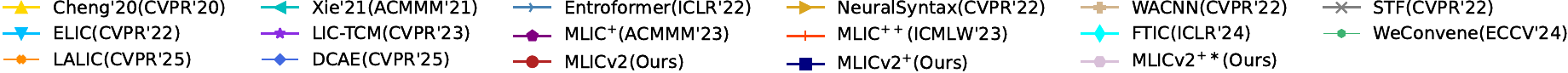}\\
    \includegraphics[scale=0.38]{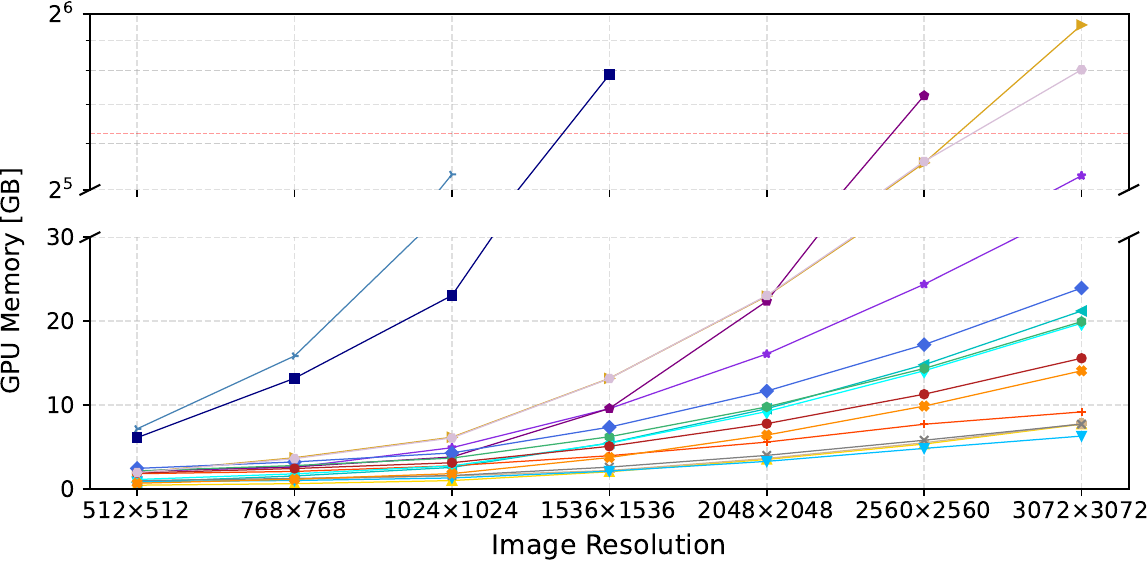}
      \includegraphics[scale=0.38]{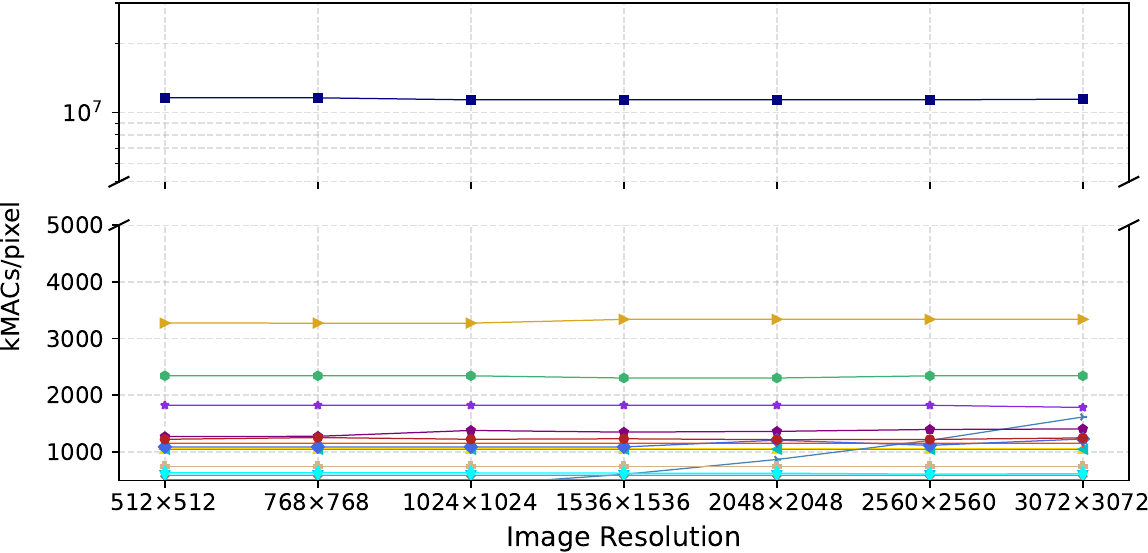}\\
    \includegraphics[scale=0.38]{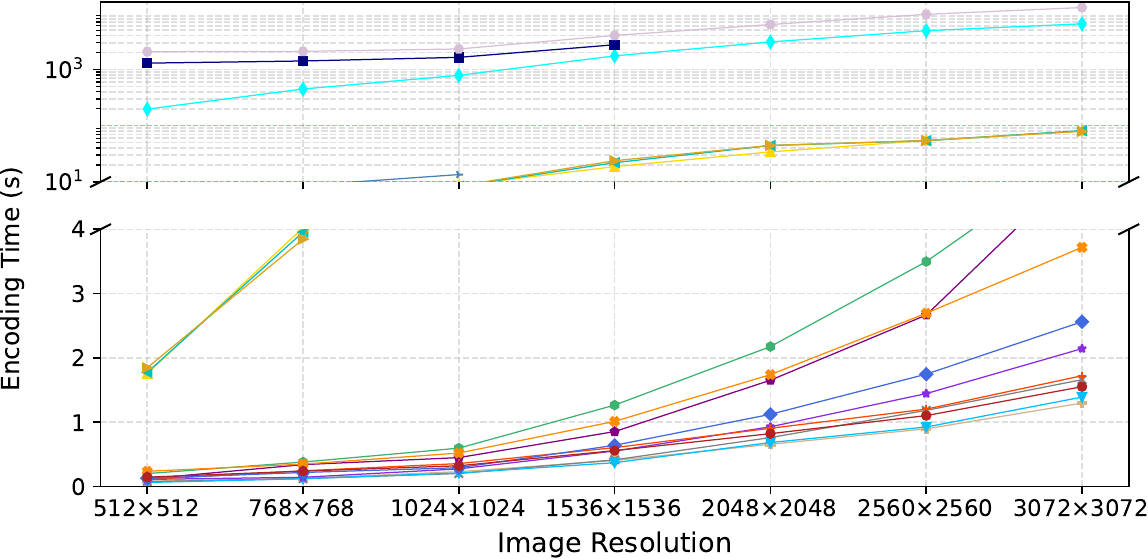}
    \includegraphics[scale=0.38]{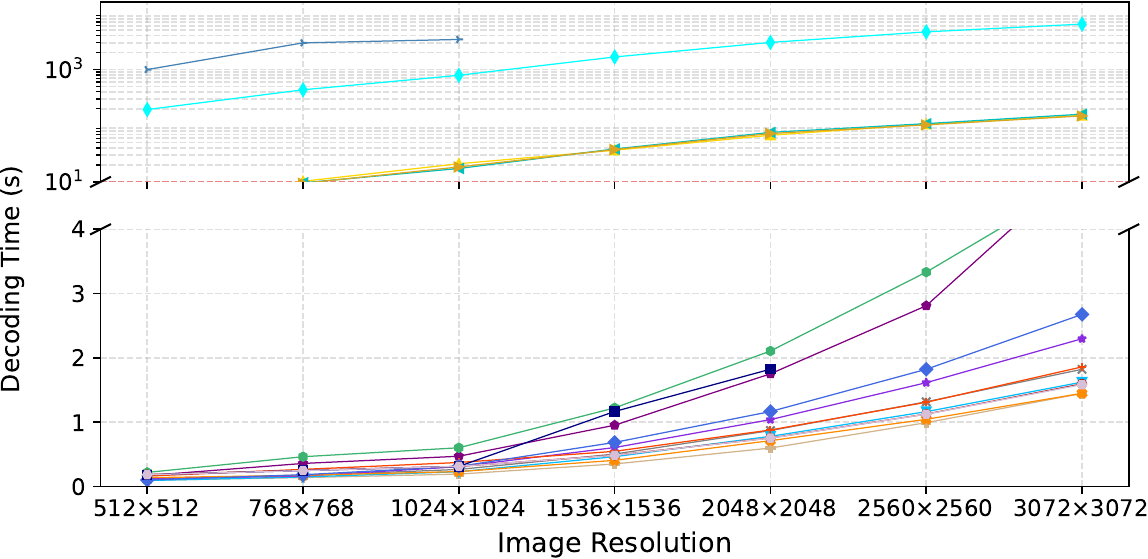}\\
  \noindent\begin{minipage}{\linewidth}
      \footnotesize
      {
      (1) In MLICv2$^{+*}$, gradient checkpointing~\cite{chen2016training} is employed during the SGA iterations.
      }
      \end{minipage}
  \noindent\begin{minipage}{\linewidth}
    \footnotesize
    {
    (2) {The MACs of MLICv2$^{+}$ are theoretically estimated based on the number of SGA iterations, 
    following the observation from Epoch AI that backward and forward passes almost always have MAC ratios close to 2:1\footnotemark[4].}
    The MACs of MLICv2$^{+*}$ are not reported due to the use of gradient checkpointing, which prevents accurate profiling.
    }
    \end{minipage}
    \noindent\begin{minipage}{\linewidth}
      \footnotesize
      {
      (3) The lack of reported results for some models at certain resolutions is due to out-of-memory (OOM) issues or inaccurate profiling.
      }  
      \end{minipage}
  \caption{{GPU memory consumption, Forward kMACs/pixel, encoding time and decoding time comparisons among our proposed models and recent learned image compression models
  ~\cite{cheng2020learned,xie2021enhanced,wang2022neural,zou2022the,he2022elic,liu2023learned,li2024frequencyaware,jiang2022mlic,jiang2023mlicpp}.
    }}\label{fig:complex}
\end{figure*}
\subsection{Rate-Distortion Performance}
\label{sec:exp:perf}
\subsubsection{Quantitive Results}
For a fair comparison with learned image compression models, VTM-17.0 Intra 
is evaluated in YUV444 color space under \textit{encoder\_intra\_vtm.cfg}.
To demonstrate the superiority of our MLICv2 and MLICv2$^{+}$, we compare with 
recent learned image compression models~\cite{li2020efficient,cheng2020learned,wu22block,
minnen2020channel,qian2020learning,xie2021enhanced,chen2021nic,guo2021causal,
qian2022entroformer,zhu2022transformerbased,zou2022the, 
he2022elic,koyuncu2022contextformer,fu2023learned,feng2023nvtc,
liu2023learned,li2024frequencyaware,jiang2024llic,fu2024weconvene,feng2025linear,lu2025learned}.\par
The rate-distortion curves are presented in Fig.~\ref{fig:rd}.
When compared with our baseline MLIC$^{++}$, MLICv2$^{+}$
performs significantly better, especially at high bit-rates
and on high-resolution images~\cite{tecnick2014TESTIMAGES,CLIC2020}.
Specifically, 
our MLICv2 achieves a maximum improvement of $0.6, 0.5, 0.5$ dB 
when the bpp is around $0.8, 0.55, 0.6$ on Kodak, Tecnick, CLIC Pro Val, respectively.
Our MLICv2$^{+}$ achieves larger improvement compared to MLIC$^{++}$~\cite{jiang2023mlicpp} due to 
adopted latent refinement~\cite{yang2020improving} for enhanced instance adaptability. 
The average improvements of latent refinement on MLICv2 on Kodak~\cite{kodak}, 
Tecnick~\cite{tecnick2014TESTIMAGES}, CLIC Pro Val~\cite{CLIC2020} are $0.22, 0.15, 0.15$ dB,
respectively.\par
\footnotetext[4]{{\texttt{\url{https://epoch.ai/blog/backward-forward-FLOP-ratio}}}}
{BD-Rate} reductions over VTM-17.0 Intra are presented in Table~\ref{tab:rd}.
Compared with existing methods, our MLICv2 and MLICv2$^{+}$ achieve
state-of-the-art performance, reducing {BD-Rates} by $16.54\%$ and $20.46\%$ on Kodak.
When compared to the baseline model MLIC$^{++}$~\cite{jiang2023mlicpp}, 
our MLICv2 and MLICv2$^+$ reduce $3.15\%$ and $7.07\%$ more bit-rates over VTM-17.0 Intra on Kodak~\cite{kodak}, respectively.
Our MLICv2 and MLICv2$^{+}$ also achieve superior performances on high-resolution datasets Tecnick~\cite{tecnick2014TESTIMAGES}
and CLIC Pro Val~\cite{CLIC2020}. Specifically, our MLICv2 and MLICv2$^{+}$ reduce {BD-Rates} by
 $21.61\%$ and $24.35\%$ on Tecnick~\cite{tecnick2014TESTIMAGES} and 
 $16.05\%$ and $19.14\%$ on CLIC Pro Val~\cite{CLIC2020}.
\subsubsection{Qualitative Results}
We also compare the qualitative performance
to MLIC$^{++}$, ELIC~\cite{he2022elic}, STF, WACNN~\cite{zou2022the},
Cheng'20~\cite{cheng2020learned} and VTM-17.0 Intra on subjective qualities.
The windowsill of reconstructions are cropped to patches for clearer comparisons.
The reconstructions of MLICv2 have sharper textures and retain
more details. In terms of visual quality, our MLICv2 has
significant improvements on rate-perception performance
compared to other models.
\begin{figure*}
\centering
\subfloat[\texttt{Kodim01}]{\includegraphics[width=.16\columnwidth]{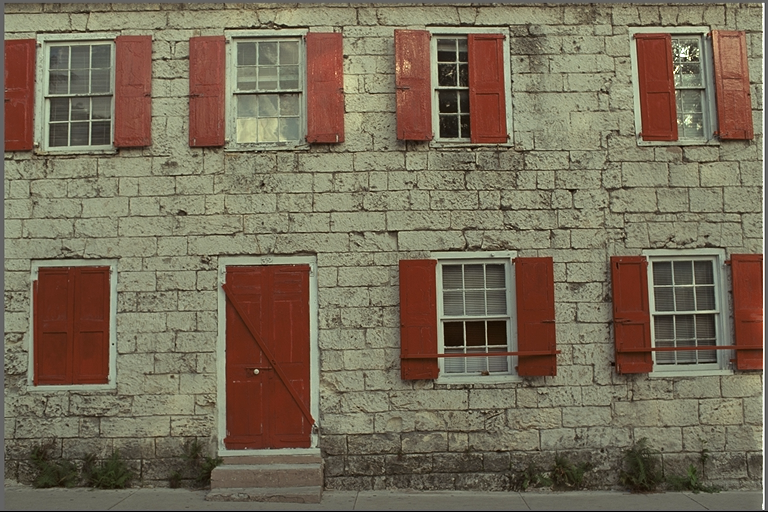}}
\subfloat[$j= 900$]{\includegraphics[width=.16\columnwidth]{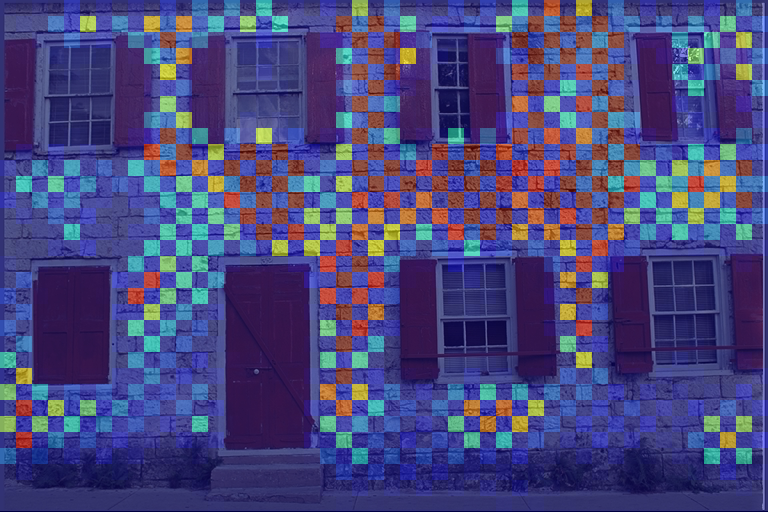}}
\subfloat[$j = 1101$]{\includegraphics[width=.16\columnwidth]{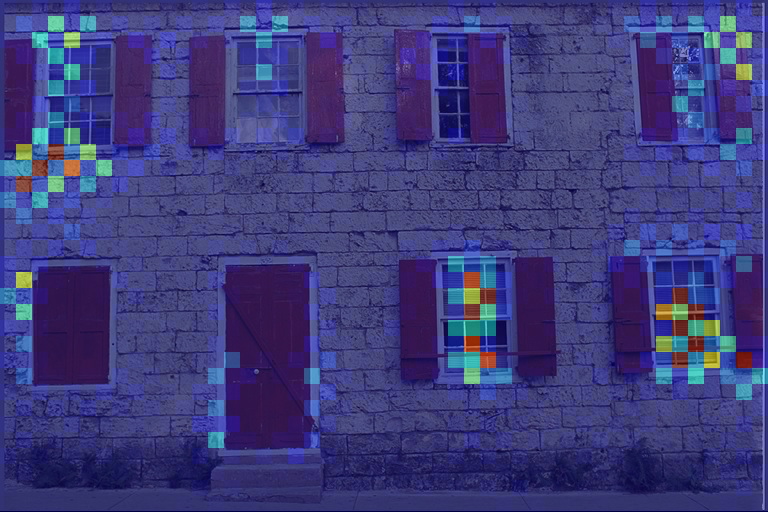}}
\quad
\subfloat[\texttt{Kodim20}]{\includegraphics[width=.16\columnwidth]{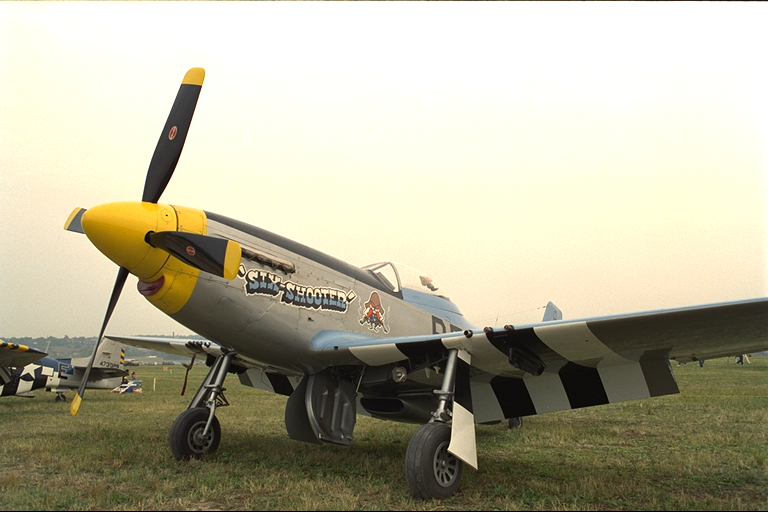}}
\subfloat[$j= 1018$]{\includegraphics[width=.16\columnwidth]{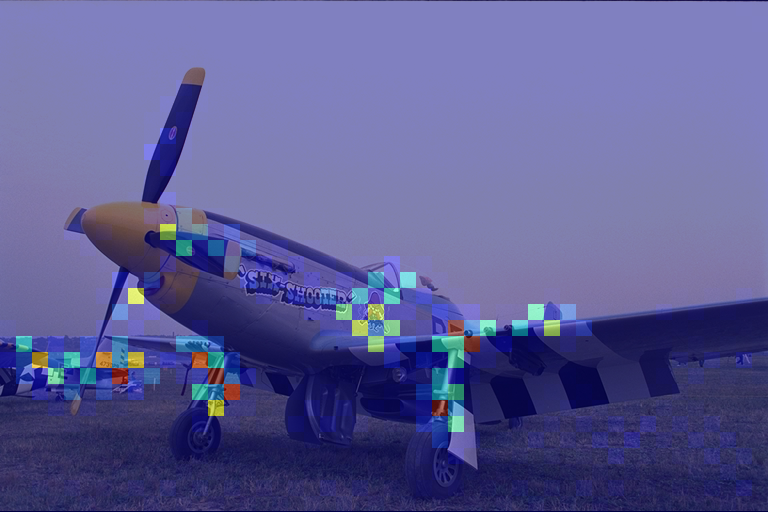}}
\subfloat[$j = 1487$]{\includegraphics[width=.16\columnwidth]{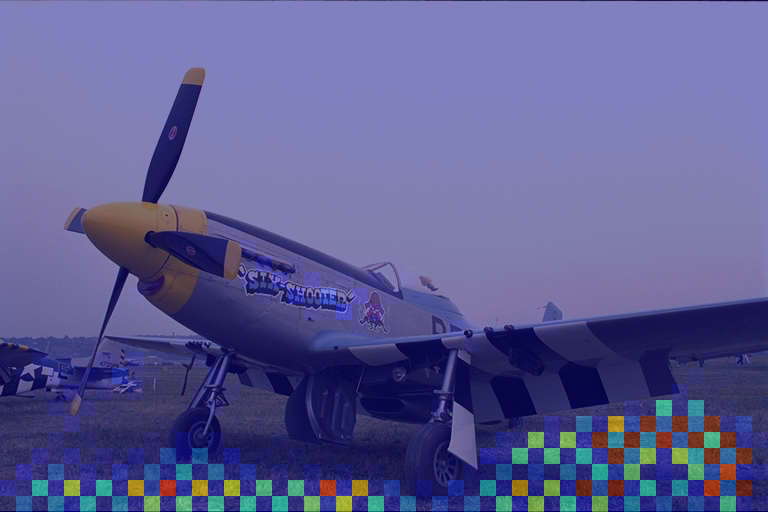}}
\caption{{Visualization of attention maps of hyperprior guided intra slice global context module. The inputs are Kodim01 and Kodim20.
The attention map is computed by $\textrm{Softmax}_2(\hat{\boldsymbol{H}}_{na,q})[j]\textrm{Softmax}_1(\hat{\boldsymbol{H}}_{ac,k})^{\top}$, where $j$ is 
the query index.}}
\label{fig:cs}
\end{figure*}
\subsection{Computational Complexity}
The model complexities and sizes are summarized in Fig.~\ref{fig:complex} and Table~\ref{tab:param}, respectively. 
{To evaluate complexity across resolutions, 
16 images from the LIU4K dataset~\cite{liu2020comprehensive} are cropped to patches of 
$\{512^2, 768^2, 1024^2, 1536^2, 2048^2, 2560^2, 3072^2\}$ pixels}, covering a wide range of practical image sizes. 
We report forward GPU memory, encoding and decoding times (including arithmetic coding), and forward {kMACs/pixel} as complexity metrics. 
Since the number of skipped elements varies with bit rate, results for MLICv2 and MLICv2$^+$ are reported at $\lambda=0.0130$.\par
\par
Due to the high memory consumption of SGA, we adopt gradient checkpointing~\cite{chen2016training} to alleviate GPU memory limitations. 
{Instead of storing all intermediate activations during the forward pass, only a small subset necessary for gradient computation is cached,
while the remaining activations are recomputed on-the-fly during backpropagation.}
This strategy substantially reduces peak memory usage at the cost of increased computation time.    
\subsubsection{On Forward GPU Memory}
Compared to MLIC$^{+}$\cite{jiang2022mlic}, MLICv2 significantly reduces memory usage thanks to its linear-complexity entropy model. 
Relative to MLIC$^{++}$\cite{jiang2023mlicpp}, it incurs slightly higher memory for high-resolution images due to the additional modules introduced in this work, but still requires less memory than Xie’21~\cite{xie2021enhanced}, Qian’21~\cite{qian2020learning}, Entroformer~\cite{qian2020learning}, and LIC-TCM~\cite{liu2023learned}. This modest increase is justified by the corresponding performance improvements. Latent refinement, while increasing memory during backpropagation, is optional and primarily benefits scenarios where encoding complexity is less critical. Since images are typically encoded once but decoded multiple times, 
the added encoding cost is acceptable given the notable gains in rate-distortion performance.
\begin{figure}
\centering
\includegraphics[width=0.5\linewidth]
{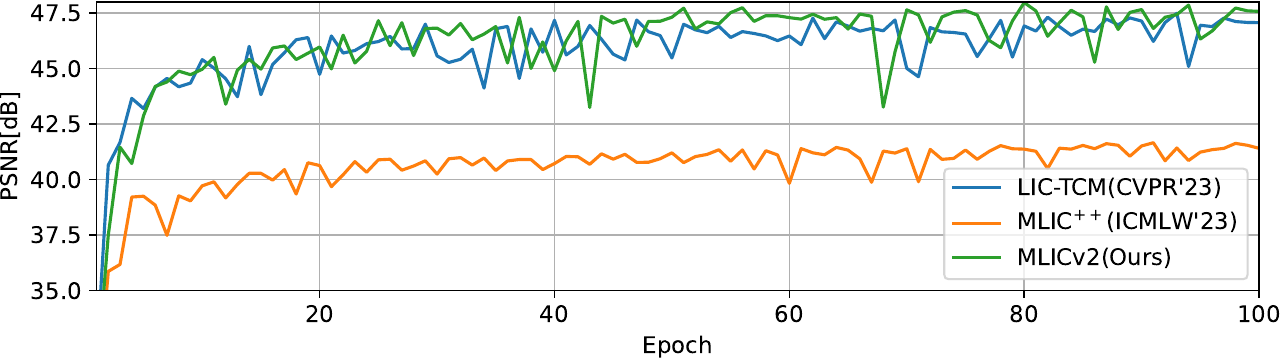}
\caption{{Upper-bound reconstruction quality (PSNR) on the Kodak dataset without entropy constraints.}}
\label{fig:bound}
\end{figure}
\begin{table*}
\centering
\footnotesize
\caption{{Comparisons among different transforms of proposed MLICv2, LIC-TCM~\cite{liu2023learned},
MLIC$^{++}$, ELIC~\cite{he2022elic}, and Cheng'20~\cite{cheng2020learned}. All method are quiped with autoregressive context module~\cite{minnen2018image}.}}
\setlength{\tabcolsep}{2mm}{
\begin{tabular}{@{}cccccccccccccc@{}}
\toprule
\multicolumn{1}{c|}{\multirow{1}{*}{Transform Architecture}}    & \multicolumn{1}{c}{Cheng'20~\cite{cheng2020learned}}& \multicolumn{1}{c}{SwinT~\cite{zhu2022transformerbased}}& \multicolumn{1}{c}{ELIC~\cite{he2022elic}}& \multicolumn{1}{c}{MLIC$^{++}$} & \multicolumn{1}{c}{LIC-TCM~\cite{liu2023learned}} & \multicolumn{1}{c}{{FTIC}~\cite{li2024frequencyaware}} & \multicolumn{1}{c}{{DCAE}~\cite{lu2025learned}} & \multicolumn{1}{c}{MLICv2}    \\\midrule
\multicolumn{1}{c|}{BD-Rate(\%)}                               &  0.00   & -4.61     & -5.19          &   -0.41           & -8.91    & {-8.36}         & {-9.17}                  & -9.44                        \\\bottomrule  
\end{tabular}}
\label{tab:transform}
\end{table*}
\begin{figure*}
\centering
\scriptsize
\renewcommand{\arraystretch}{1}
\setlength{\tabcolsep}{1pt}

\begin{tabular}{
  >{\centering\arraybackslash}p{0.18\linewidth}
  >{\centering\arraybackslash}p{0.18\linewidth}
  >{\centering\arraybackslash}p{0.18\linewidth}
  >{\centering\arraybackslash}p{0.18\linewidth}
  >{\centering\arraybackslash}p{0.18\linewidth}
  }
\toprule
\multicolumn{1}{c|}{\centering{Input Image}} &
\multicolumn{4}{c}{{Selected Channel Features of \textit{Intra-Slice Local} Context and Mean Attention Weights}} \\
\midrule
{\scriptsize Kodak: \texttt{Kodim20}} &
{Mean Attention Weight: 0.1555} &
{Mean Attention Weight: 0.1211} &
{Mean Attention Weight: 0.0553} &
{Mean Attention Weight: 0.0127} \\
\includegraphics[width=0.8\linewidth]{figures/kodim20.png} &
\includegraphics[width=0.8\linewidth]{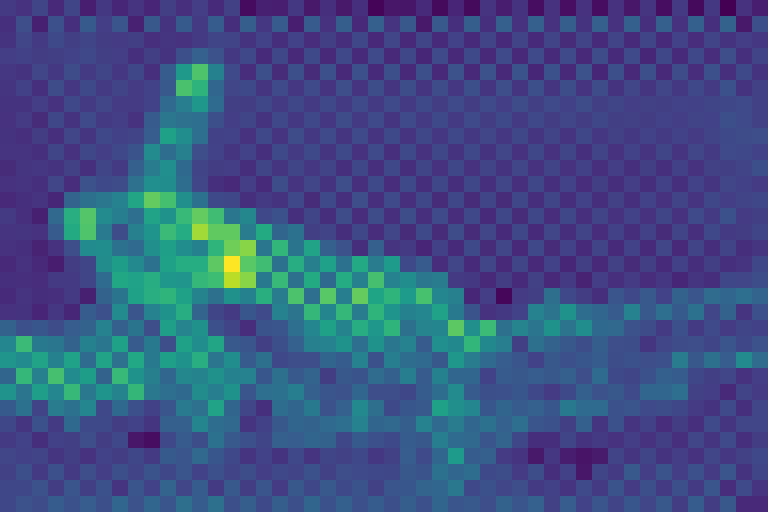} & 
\includegraphics[width=0.8\linewidth]{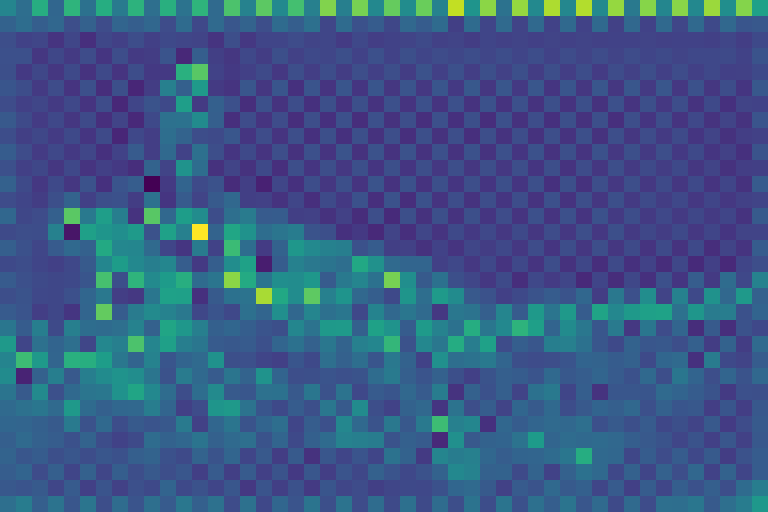} & 
\includegraphics[width=0.8\linewidth]{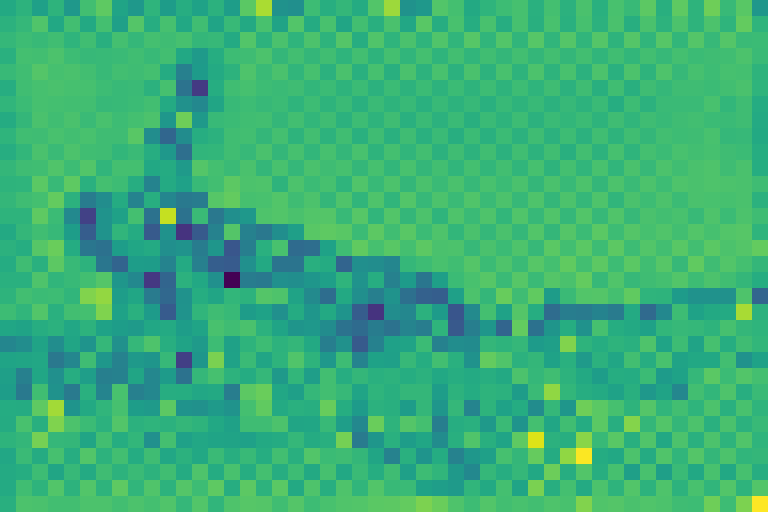} & 
\includegraphics[width=0.8\linewidth]{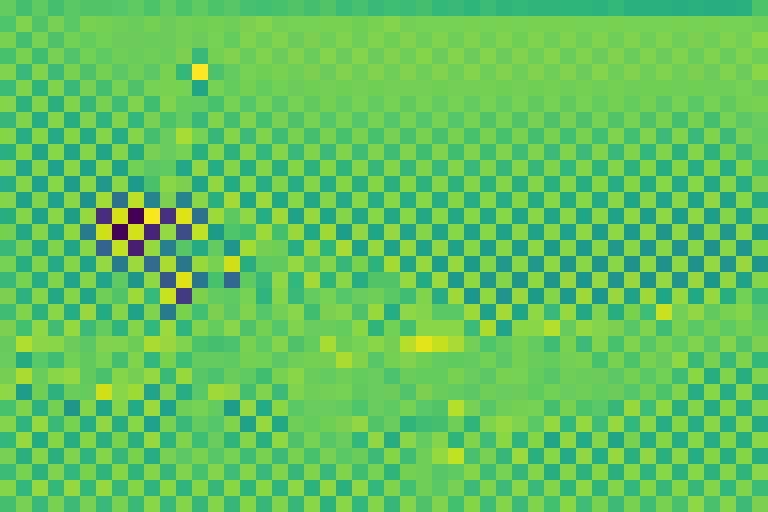} \\
{\scriptsize Kodak: \texttt{Kodim21}} &
{Mean Attention Weight: 0.1471} &
{Mean Attention Weight: 0.0780} &
{Mean Attention Weight: 0.0401} &
{Mean Attention Weight: 0.0184} \\
\includegraphics[width=0.8\linewidth]{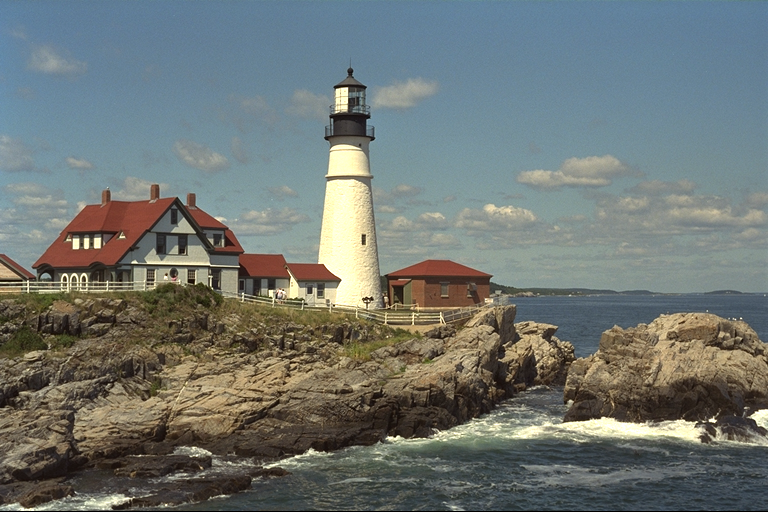} &
\includegraphics[width=0.8\linewidth]{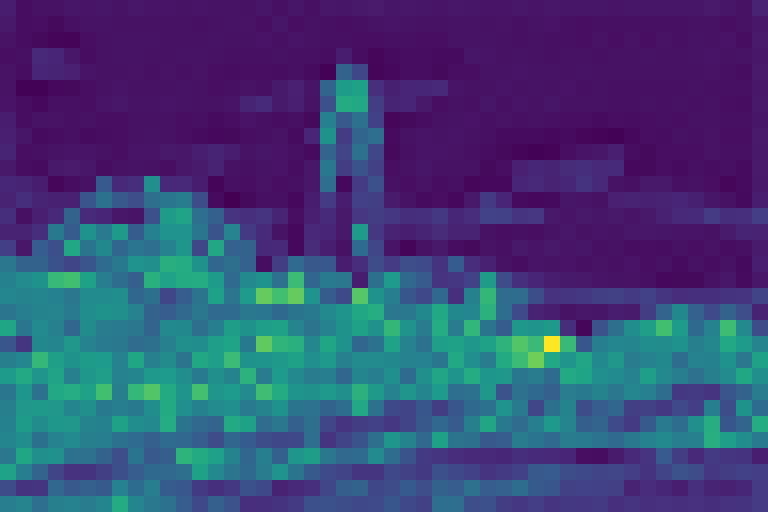} &
\includegraphics[width=0.8\linewidth]{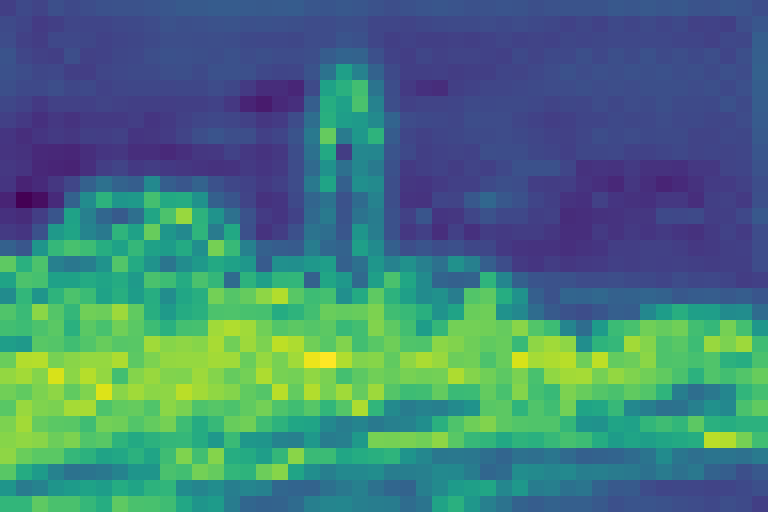} &
\includegraphics[width=0.8\linewidth]{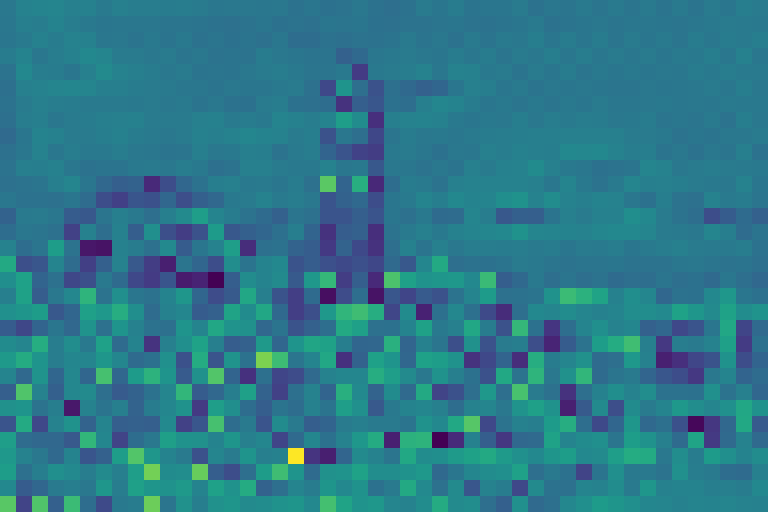} &
\includegraphics[width=0.8\linewidth]{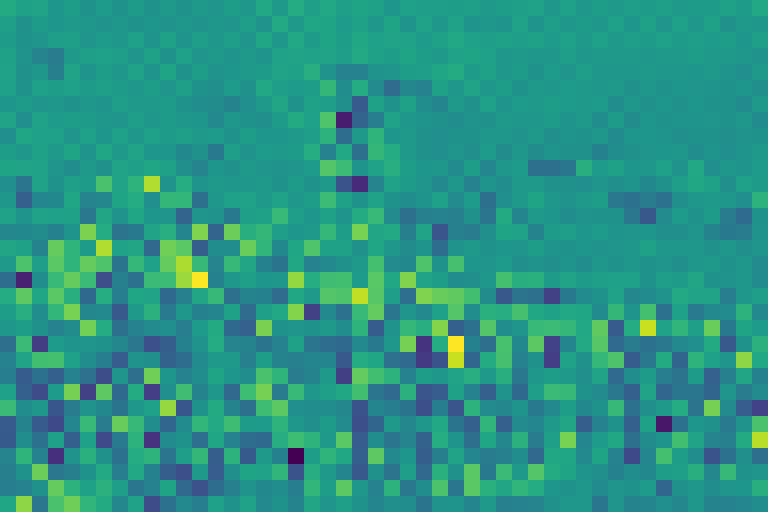} \\
\midrule
\multicolumn{1}{c|}{\centering{Input Image}} &
\multicolumn{4}{c}{{Selected Channel Features of \textit{Inter-Slice Local} Context and Mean Attention Weights}} \\
\midrule
{\scriptsize Kodak: \texttt{Kodim20}} &
{Mean Attention Weight: 0.0933} &
{Mean Attention Weight: 0.0822} &
{Mean Attention Weight: 0.0503} &
{Mean Attention Weight: 0.0361} \\
\includegraphics[width=0.8\linewidth]{figures/kodim20.png} &
\includegraphics[width=0.8\linewidth]{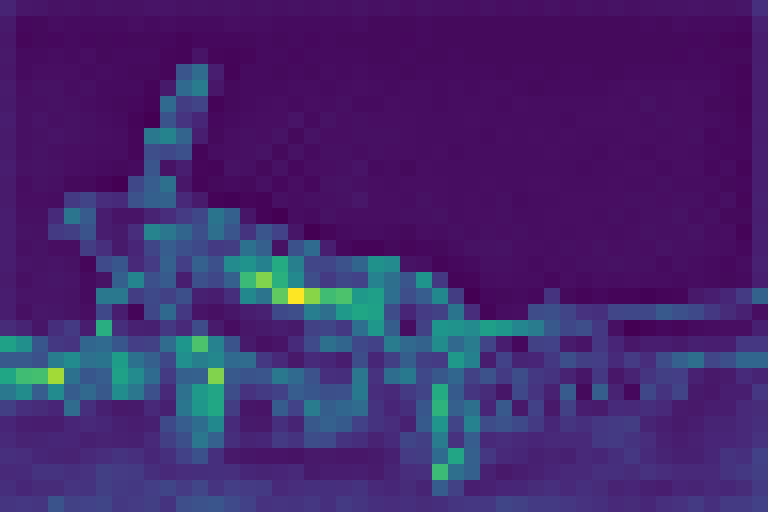} &
\includegraphics[width=0.8\linewidth]{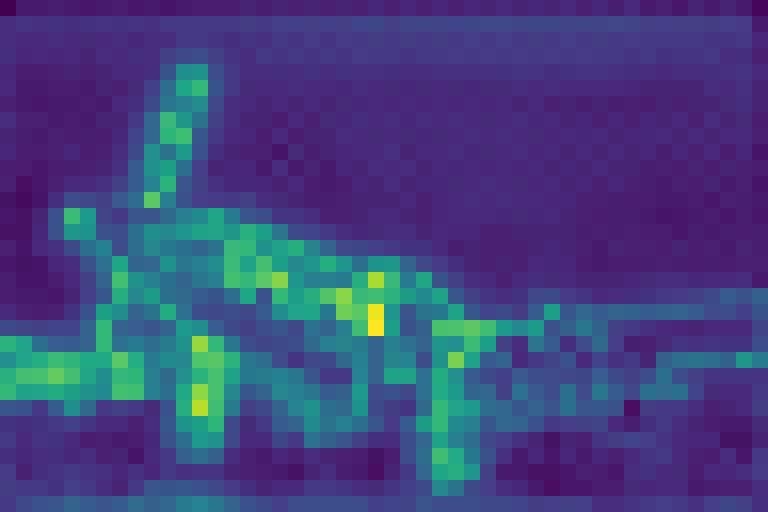} &
\includegraphics[width=0.8\linewidth]{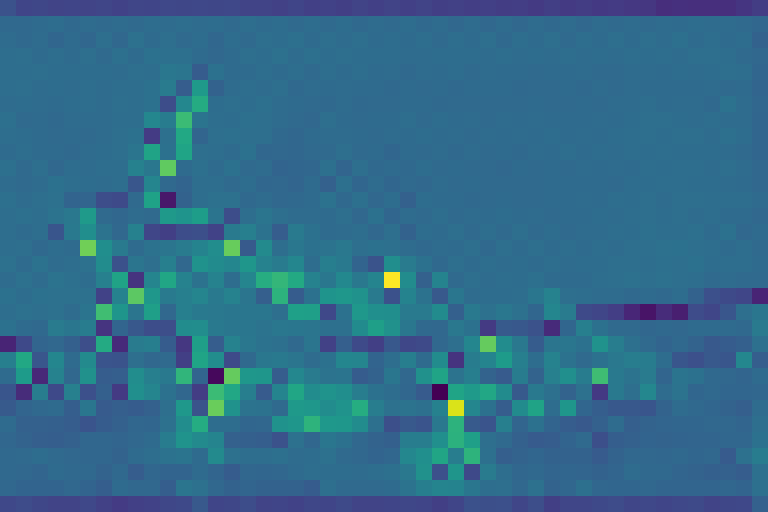} &
\includegraphics[width=0.8\linewidth]{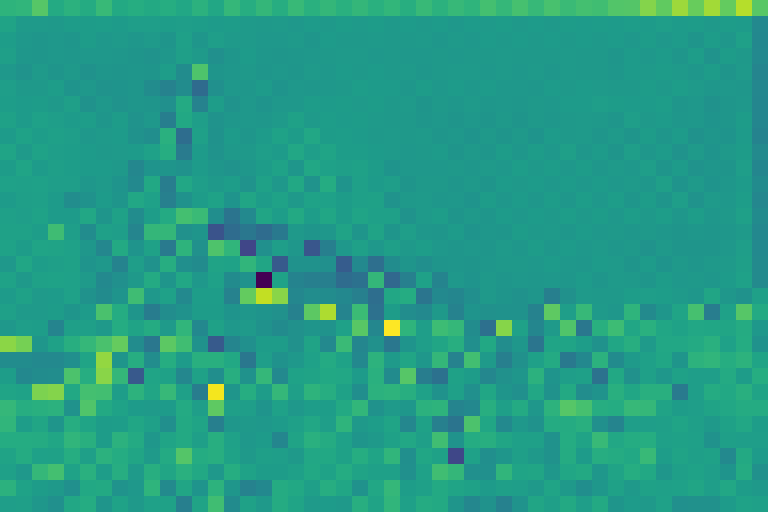} \\
{\scriptsize Kodak: \texttt{Kodim21}} &
{Mean Attention Weight: 0.0909} &
{Mean Attention Weight: 0.0720} &
{Mean Attention Weight: 0.0452} &
{Mean Attention Weight: 0.0327} \\
\includegraphics[width=0.8\linewidth]{figures/kodim21.png} &
\includegraphics[width=0.8\linewidth]{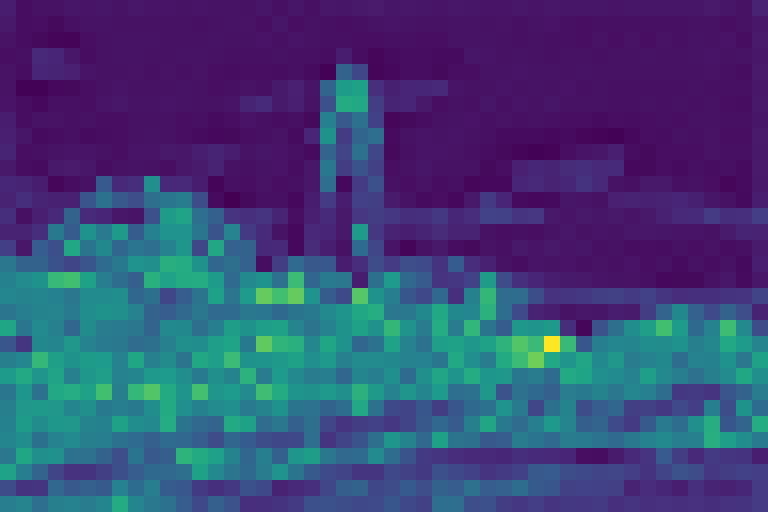} &
\includegraphics[width=0.8\linewidth]{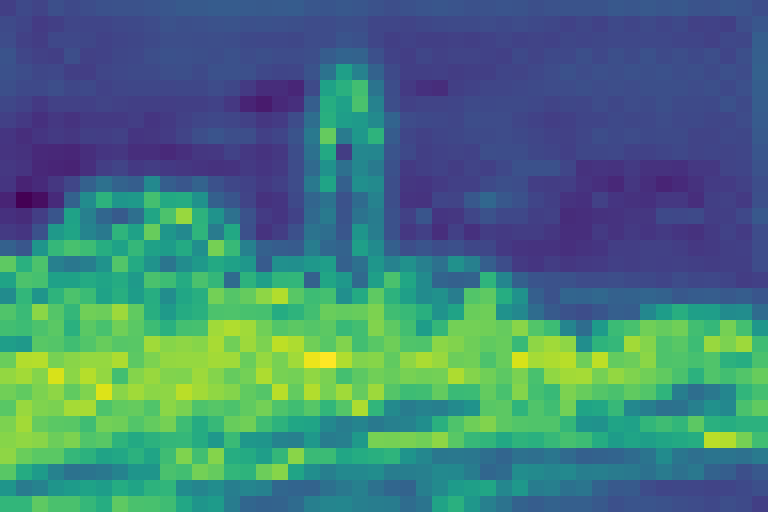} &
\includegraphics[width=0.8\linewidth]{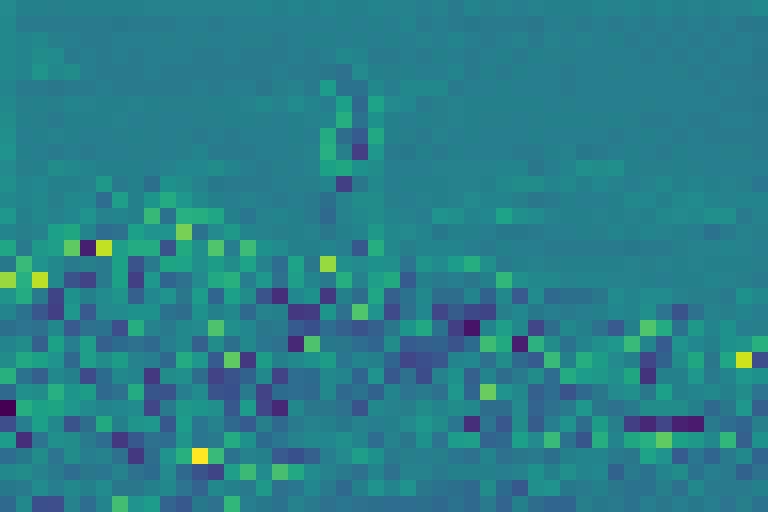} &
\includegraphics[width=0.8\linewidth]{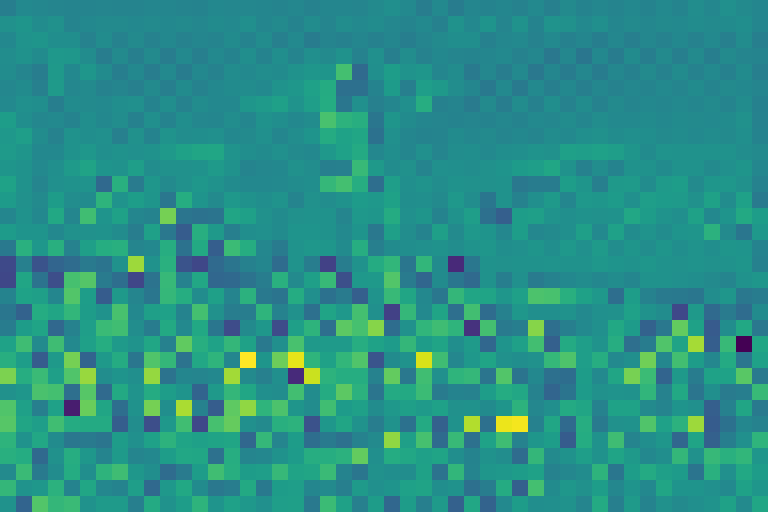} \\
\midrule
\multicolumn{1}{c|}{\centering{Input Image}} &
\multicolumn{4}{c}{{Selected Channel Features of \textit{Intra-Slice Global} Context and Mean Attention Weights}} \\
\midrule
{\scriptsize Kodak: \texttt{Kodim20}} &
{Mean Attention Weight: 0.0728} &
{Mean Attention Weight: 0.0704} &
{Mean Attention Weight: 0.0619} &
{Mean Attention Weight: 0.0393} \\
\includegraphics[width=0.8\linewidth]{figures/kodim20.png} &
\includegraphics[width=0.8\linewidth]{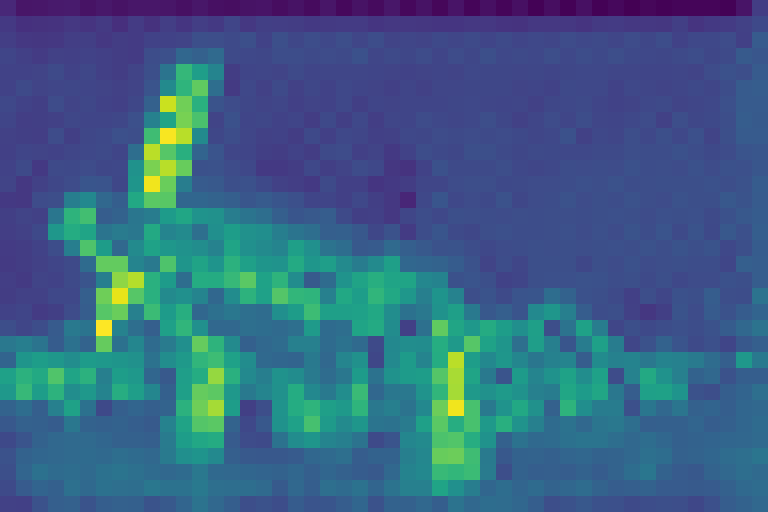} &
\includegraphics[width=0.8\linewidth]{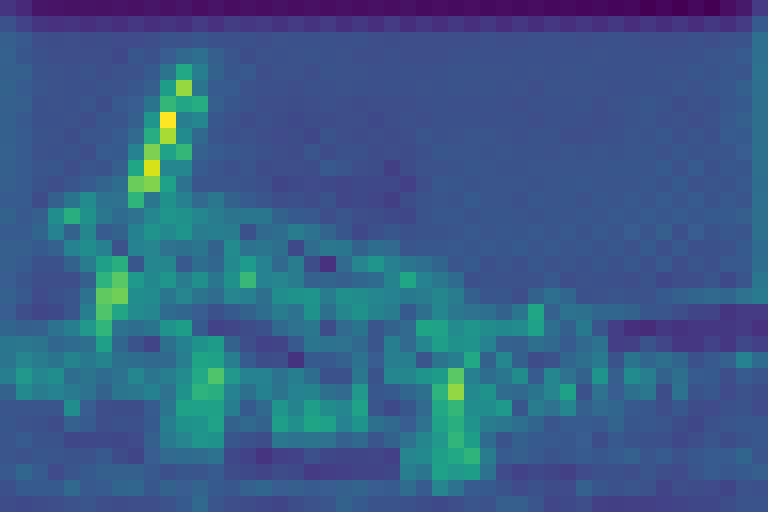} &
\includegraphics[width=0.8\linewidth]{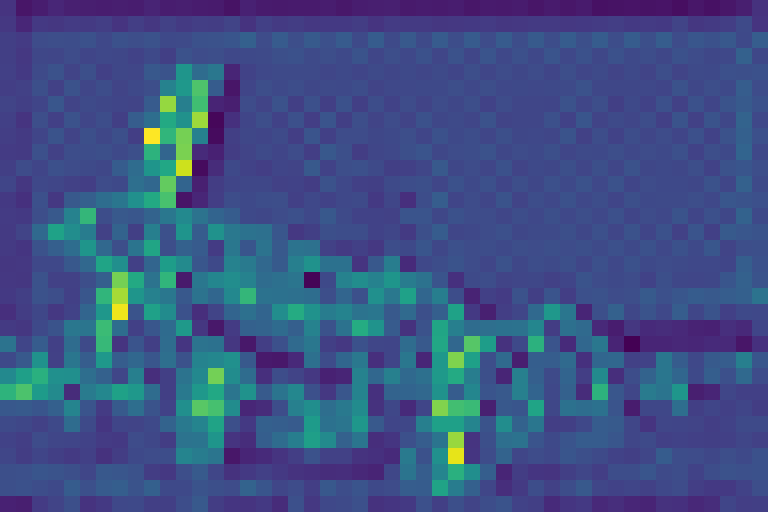} &
\includegraphics[width=0.8\linewidth]{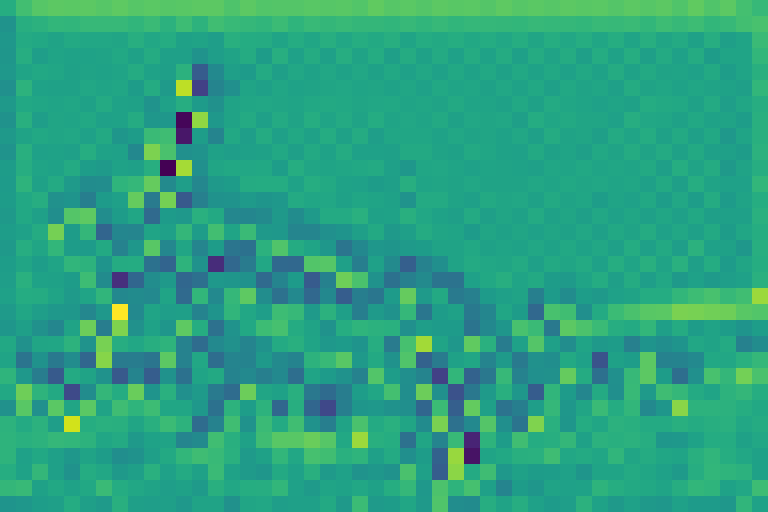} \\
{\scriptsize Kodak: \texttt{Kodim21}} &
{Mean Attention Weight: 0.0648} &
{Mean Attention Weight: 0.0623} &
{Mean Attention Weight: 0.0607} &
{Mean Attention Weight: 0.0234} \\
\includegraphics[width=0.8\linewidth]{figures/kodim21.png} &
\includegraphics[width=0.8\linewidth]{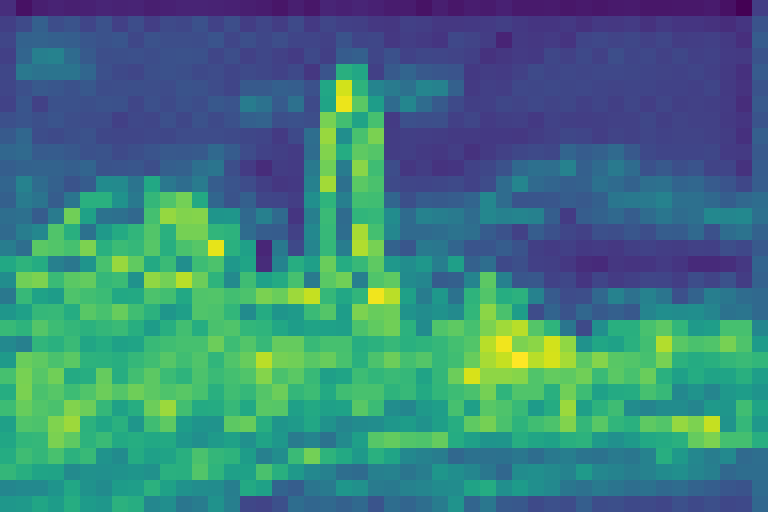} &
\includegraphics[width=0.8\linewidth]{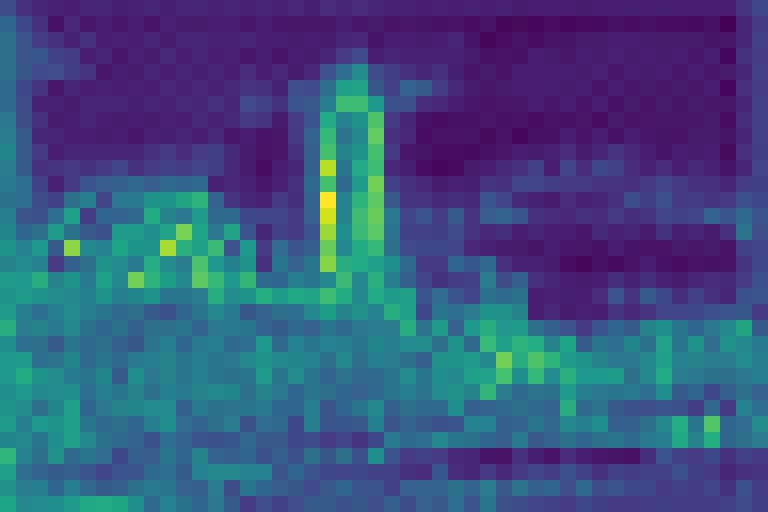} &
\includegraphics[width=0.8\linewidth]{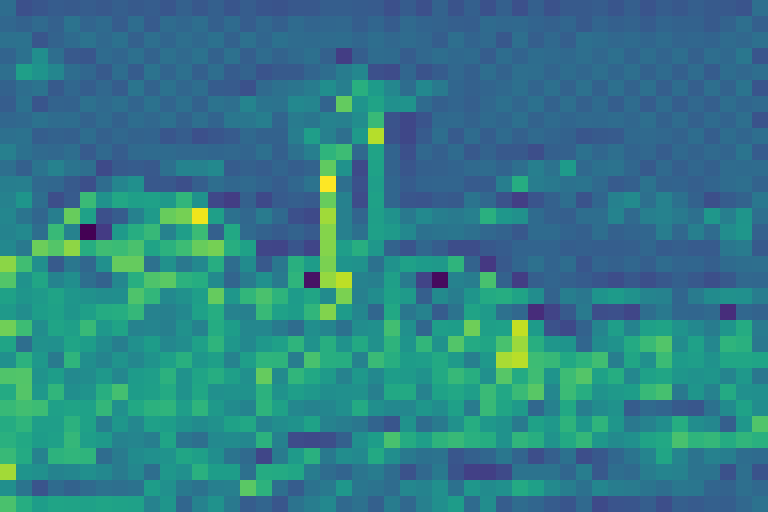} &
\includegraphics[width=0.8\linewidth]{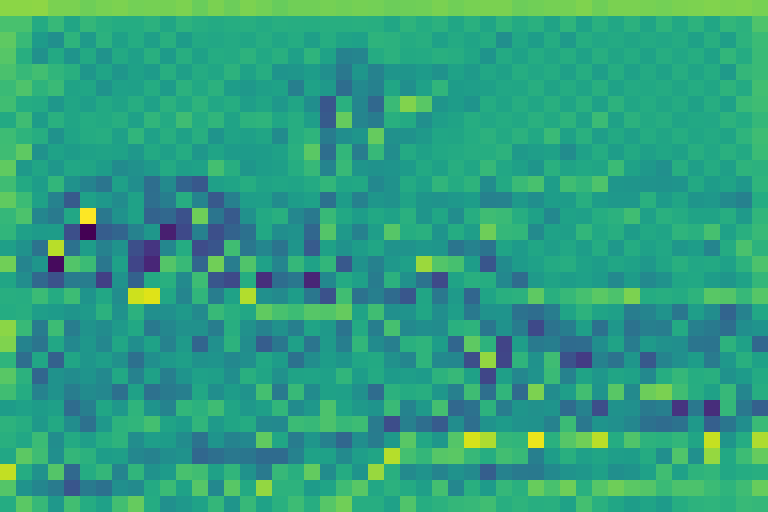} \\
\midrule
\multicolumn{1}{c|}{\centering{Input Image}} &
\multicolumn{4}{c}{{Selected Channel Features of \textit{Inter-Slice Global} Context and Mean Attention Weights}} \\
\midrule
{\scriptsize Kodak: \texttt{Kodim20}} &
{Mean Attention Weight: 0.1424} &
{Mean Attention Weight: 0.1378} &
{Mean Attention Weight: 0.0783} &
{Mean Attention Weight: 0.0134} \\
\includegraphics[width=0.8\linewidth]{figures/kodim20.png} &
\includegraphics[width=0.8\linewidth]{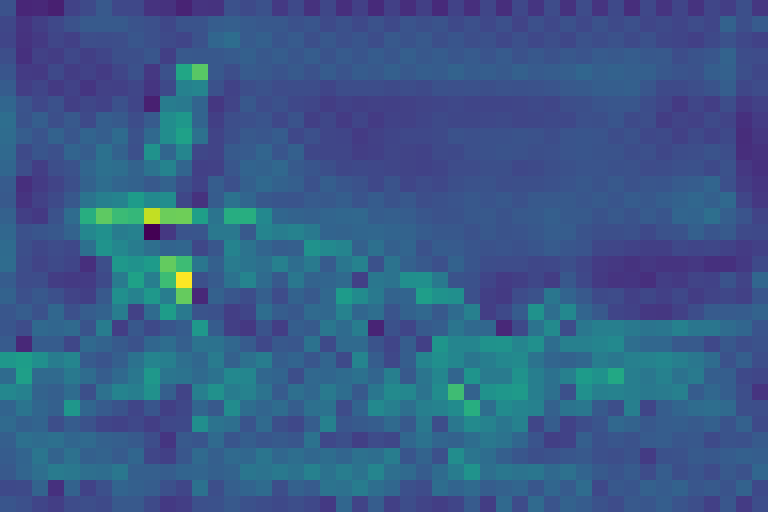} &
\includegraphics[width=0.8\linewidth]{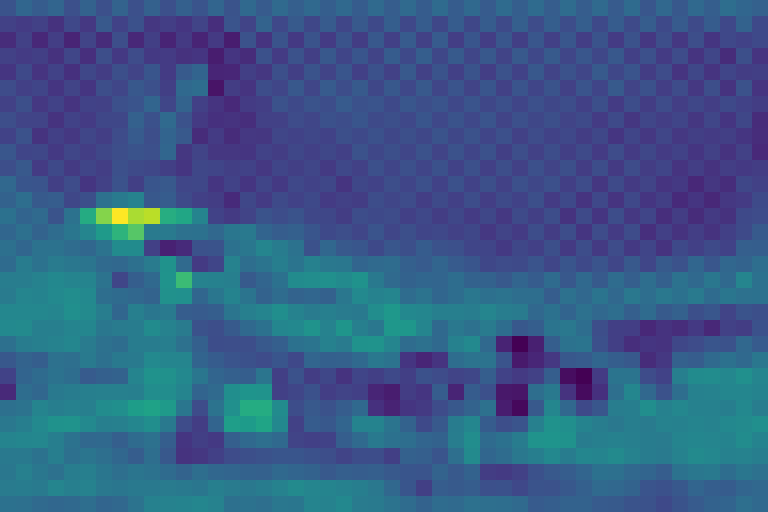} &
\includegraphics[width=0.8\linewidth]{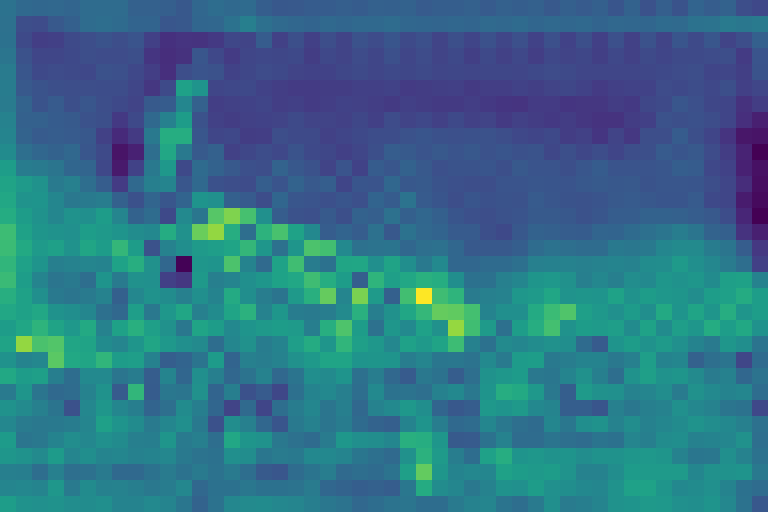} &
\includegraphics[width=0.8\linewidth]{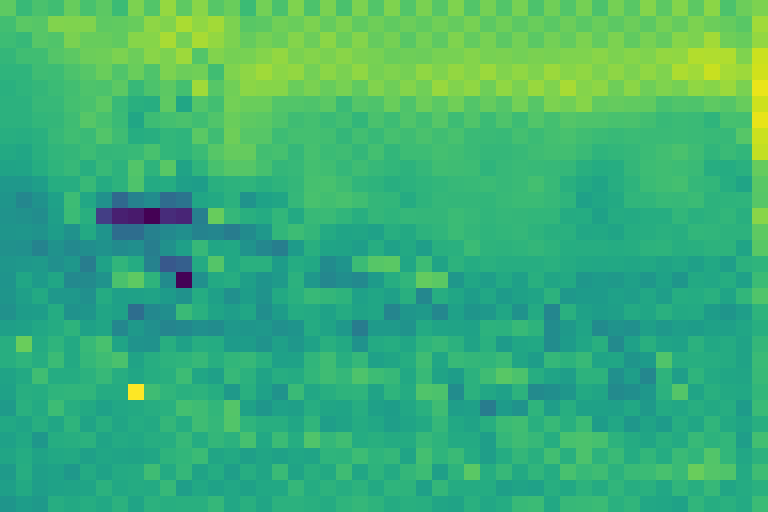} \\
{\scriptsize Kodak: \texttt{Kodim21}} &
{Mean Attention Weight: 0.2302} &
{Mean Attention Weight: 0.1499} &
{Mean Attention Weight: 0.1391} &
{Mean Attention Weight: 0.0107} \\
\includegraphics[width=0.8\linewidth]{figures/kodim21.png} &
\includegraphics[width=0.8\linewidth]{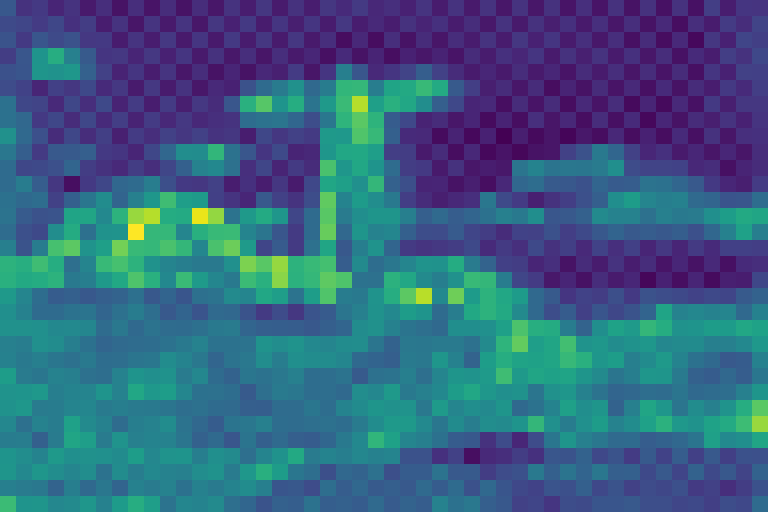} &
\includegraphics[width=0.8\linewidth]{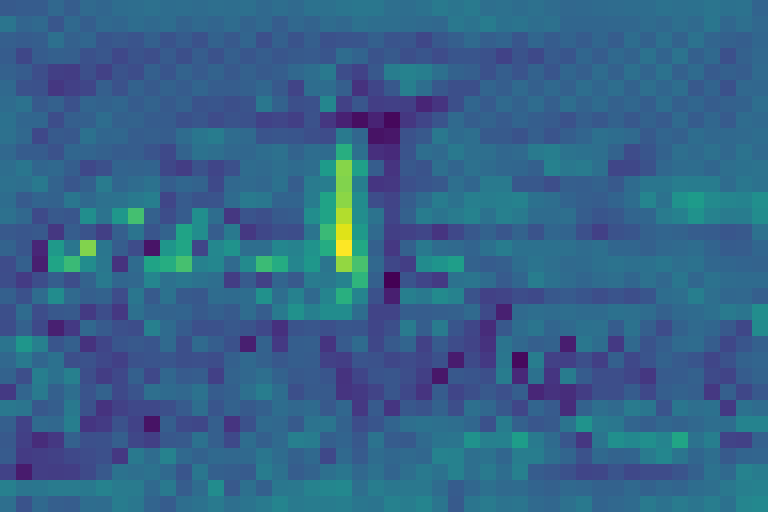} &
\includegraphics[width=0.8\linewidth]{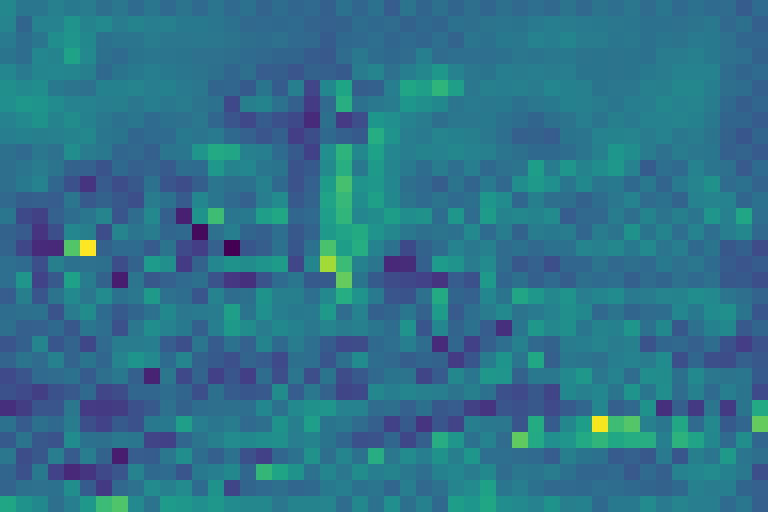} &
\includegraphics[width=0.8\linewidth]{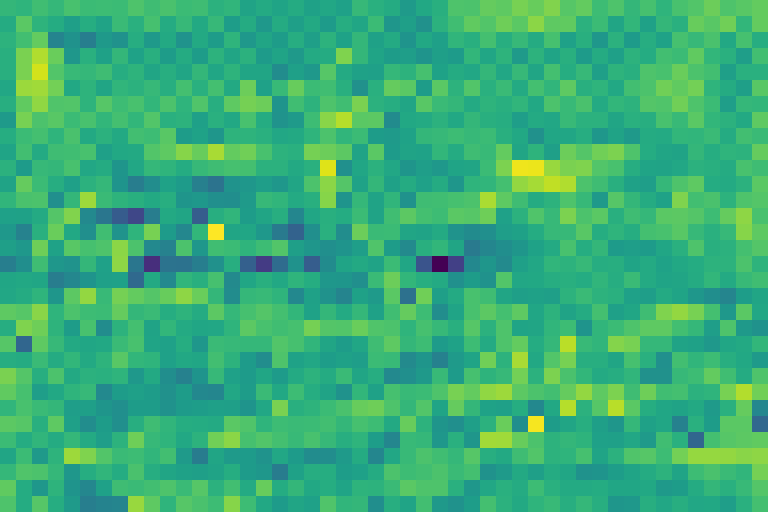} \\
\bottomrule
\end{tabular}

\caption{{Visualization of selected channel features and their corresponding Mean Attention Weights.
All channels are sampled from the context features (before context reweighting) of the last slice.
The visualized feature maps are normalized to the range [0, 1].}}
\label{fig:visualcomp2}
\end{figure*}
\begin{table*}
  \centering
  \footnotesize
  \caption{{MACs contribution of the introduced modules to the entropy model on the Kodak dataset}}
  \setlength{\tabcolsep}{2mm}{
  \begin{tabular}{@{}cccccccccccccc@{}}
  \toprule
  \multicolumn{1}{c|}{\multirow{1}{*}{Context Module}}    & \multicolumn{1}{c}{MLIC$^{++}$~\cite{jiang2023mlicpp}}& \multicolumn{1}{c}{MLICv2}& \multicolumn{1}{c}{{HGCP}}& \multicolumn{1}{c}{2D RoPE} & \multicolumn{1}{c}{{CR}}   & \multicolumn{1}{c}{{GSC}}    \\\midrule
  \multicolumn{1}{c|}{GMACs}                               &  {$508.49$}   & {{$506.76$}}     & {{$1.62$}}          &   {{$2.4\times 10^{-5}$}}           & {{$3.46$}}                               & {{$1.73$}}                         \\\bottomrule  
\end{tabular}}
  \label{tab:context_complex}
\end{table*}
\subsubsection{On Forward {kMACs/pixel}}
Compared to MLIC$^{+}$, LIC-TCM~\cite{liu2023learned}, our MLICv2 has lower {kMACs/pixel}.
Compared to MLIC$^{++}$, our MLICv2 exhibits slightly higher {kMACs/pixel}.
This increase can be attributed to two primary factors: 
the proposed channel reweighting module and selective compression module. 
However, considering the significant performance gains achieved through these enhancements, 
the marginal increase in MACs is deemed acceptable. 
The trade-off between computational complexity and improved compression efficiency 
is well-balanced in our proposed model.
\subsubsection{On Encoding Time and Decoding Time}
Compared to MLIC and MLIC$^{+}$~\cite{jiang2022mlic}, 
MLICv2 achieves higher encoding and decoding efficiency due to its linear-complexity components. 
Relative to MLIC$^{++}$\cite{jiang2023mlicpp}, it further reduces encoding and decoding time by efficiently skipping many zero elements, 
which is particularly beneficial for high-resolution images where zeros are more frequent. Latent refinement, while involving multiple backpropagation steps and increasing encoding time, 
is optional and suited for scenarios where encoding complexity is less critical. 
Since images are typically encoded once but decoded multiple times, 
the extra encoding cost is justified by the performance gains. 
Notably, MLICv2$^{+}$ maintains decoding complexity nearly identical to MLICv2, 
with both substantially faster than MLIC$^{++}$, recent WeConvene~\cite{fu2024weconvene} and DCAE~\cite{lu2025learned}.
\subsubsection{On Model Size}
{
The model size comparisons are presented in Table~\ref{tab:param}.
Although MLICv2 incorporates several additional components, 
its parameter count (84.3M) remains comparable to that of MLIC$^{++}$ (83.5M). 
The increase in parameters is minimal, primarily because we deliberately reduce the 
number of channels for intermediate features in the context modules of MLICv2 to control 
overall model size. }
\begin{table*}
\scriptsize
\centering
\renewcommand\arraystretch{0.95}
\caption{Complete ablation study on Kodak dataset. BD-Rate (\%) is employed to evaluate component contributions. MLICv2$^{+}$ is the anchor. Cases 1–7 show cumulative benefits, Cases 9–15 show individual contributions. The original Cases 8 and 16 are merged as a separate baseline.}
\setlength{\tabcolsep}{0.6mm}{
\begin{tabular}{@{}c|cccccccccccccccc@{}}
\toprule
\multicolumn{1}{c|}{\multirow{2}{*}{Context Modules}} & 
\multicolumn{1}{c}{MLICv2$^+$} & 
\multicolumn{7}{c}{Cumulative Benefits} & 
\multicolumn{7}{c}{Individual Contributions} & 
\multicolumn{1}{c}{{Baseline}} \\
\cmidrule(lr){2-2} \cmidrule(lr){3-9} \cmidrule(lr){10-16} \cmidrule(lr){17-17}
& Full & Case 1 & Case 2 & Case 3 & Case 4 & Case 5 & Case 6 & Case 7 & Case 8 & Case 9 & Case 10 & Case 11 & Case 12 & Case 13 & Case 14 & {Case 15} \\
\midrule
{ILR} & \checkmark & \checkmark & & & & & & & \checkmark & & & & & & & \\
\midrule
{GSC} & \checkmark & & \checkmark & & & & & & & \checkmark & & & & & & \\
\midrule
2D RoPE & \checkmark & \checkmark & \checkmark & \checkmark & & & & & & & \checkmark & & & & &  \\
\midrule
{CR} & \checkmark & \checkmark & \checkmark & \checkmark & \checkmark & & & & & & & \checkmark & & & &  \\
\midrule
{HGCP} & \checkmark & \checkmark & \checkmark & \checkmark & \checkmark & \checkmark & & & & & & & \checkmark & & & \\
\midrule
{STMT} & \checkmark & \checkmark & \checkmark & \checkmark & \checkmark & \checkmark & \checkmark & & & & & & & \checkmark & & \\
\midrule
Gate & \checkmark & \checkmark & \checkmark & \checkmark & \checkmark & \checkmark & \checkmark & \checkmark & & & & & & & \checkmark &  \\
\midrule
{BD-Rate (\%)} & 0.00 & 0.17 & 4.89 & 5.21 & 5.77 & 7.79 & 8.29 & 9.58 & 4.64 & 9.52 & 9.12 & 7.64 & 8.77 & 8.51 & 9.58 & 9.87 \\
\bottomrule
\end{tabular}}
\label{tab:complete_ablation}
\end{table*}
\subsection{Ablation Studies}
\label{sec:cw}
\subsubsection{Settings}
\begin{figure}
  \centering
    \includegraphics[scale=0.37]{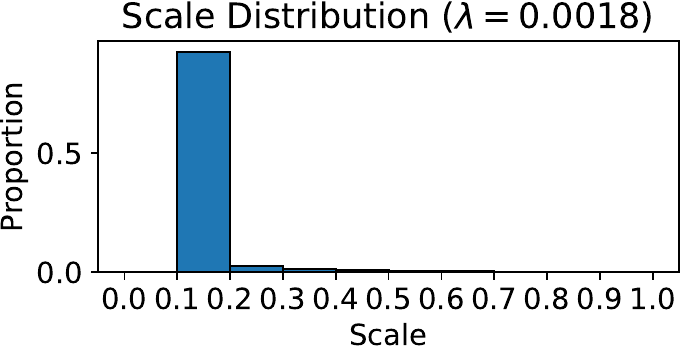}
    \includegraphics[scale=0.37]{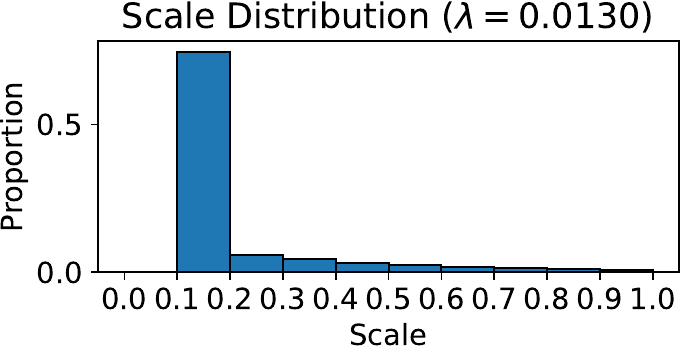}
    \caption{{Distribution of scale values in MLICv2 on Kodak~\cite{kodak}.}}
    \label{fig:scale_disrtibution}
\end{figure}
Each ablation cases is optimized for MSE using our established training strategy, with MLICv2$^+$ serving as the anchor.
We evaluate both cumulative benefits and individual contributions of each proposed component.
The results are presented in Table~\ref{tab:complete_ablation}.
\subsubsection{Analysis of Enhanced Transform}
{To demonstrate the superior capacity of our proposed transform, 
we compare the upper bounds of reconstruction quality achieved by the transforms in MLIC$^{++}$\cite{jiang2023mlicpp}, 
LIC-TCM\cite{liu2023learned}, and our MLICv2. The results are shown in Fig.~\ref{fig:bound}.
Our MetaFormer-based design achieves better reconstruction quality than MLIC$^{++}$ and 
performs comparably to LIC-TCM, highlighting its stronger representational power despite 
relying primarily on convolutional operations.}
\par
To demonstrate the effectiveness of the proposed simple token mixing block, 
we compare the transform module with the transform modules of Cheng'20~\cite{cheng2020learned}, SwinT-Charm~\cite{zhu2022transformerbased},
ELIC~\cite{he2022elic}, and LIC-TCM~\cite{liu2023learned} on rate-distortion performance.
The entropy model is aligned with spatial autoregressive entropy model~\cite{minnen2018joint}.
The results are presented in Table.~\ref{tab:transform}. Our proposed transform achieves
better performance than LIC-TCM while with about half forward MACs.
When equiped with proposed transform, the model performance gets further enhanced.
Since our baseline performance is very strong, 
the boost from the transform is slightly small relative to Table.~\ref{tab:transform}.
\subsubsection{Analysis of Enhanced Entropy Modeling}
{As shown in Table~\ref{tab:complete_ablation}, 
disabling the selective compression module results in a 0.17\% performance drop,
which is acceptable, as the skipped elements—primarily zeros—typically have very low entropy 
despite accounting for the majority of all latent elements. By omitting the encoding of these zero elements, 
we can significantly accelerate both arithmetic encoding and decoding.}
To demonstrate the effectiveness of proposed guided selective compression on acceleration, the 
skip ratio, encoding acceleration and decoding acceleration on
Kodak~\cite{kodak} and CLIC Pro Val~\cite{CLIC2020} are presented in Table~\ref{tab:skip_combined}.
{
As image quality improves, the proportion of zero elements that can be skipped decreases, 
which in turn reduces the acceleration gains for both encoding and decoding. However, 
high-resolution images usually contain more redundancy, allowing a larger number of elements to be skipped. 
This leads to greater potential for acceleration in such cases.
In addition,
we found that setting $\xi$ to 0.2, 0.3, or 0.4 has little effect on overall performance. 
This is mainly due to two reasons. {First, as illustrated in Figure~\ref{fig:scale_disrtibution}, 
most scale values fall within the range [0.1, 0.2], 
with only a small fraction between 0.2 and 0.4. Thus, using $\xi$ within this range is already effective in 
skipping most zero elements with low prediction error. The selective compression module only needs to correct 
a few wrongly skipped important elements. Second, we apply a weighted binary classification loss to train 
the selective compression module, 
using the ground-truth encoding mask as supervision.}}
\par
To demonstrate the effectiveness of hyperprior-guided global context for the first slice, the attention maps are illustrated in 
Fig.~\ref{fig:cs}, which are generally accurate, thereby enhancing the performance. 
For slices $\hat{\boldsymbol{y}}^{\geq 1}$, the previous one slice
is still employed to predict the global correlations due to the higher bit-rate of each slice compared to the side information.\par
To demonstrate the relationship between context reweighting map and image content, we visualize
the Mean Attention Weight in Fig.~\ref{fig:visualcomp2}.
\begin{table*}
\centering
\footnotesize
\caption{Model Size Comparison.}
\setlength{\tabcolsep}{0.1mm}{
\begin{tabular}{@{}c|cccccccccccc@{}}
\toprule
\multicolumn{1}{c|}{Model} & 
\rotatebox[origin=c]{45}{Xie'21~\cite{xie2021enhanced}} & 
\rotatebox[origin=c]{45}{ELIC~\cite{he2022elic}} &
\rotatebox[origin=c]{45}{WACNN~\cite{zou2022the}} & 
\rotatebox[origin=c]{45}{STF~\cite{zou2022the}} & 
\rotatebox[origin=c]{45}{LIC-TCM~\cite{liu2023learned}} &
\rotatebox[origin=c]{45}{FTIC~\cite{li2024frequencyaware}} &
\rotatebox[origin=c]{45}{MLIC$^+$~\cite{jiang2022mlic}} &
\rotatebox[origin=c]{45}{MLIC$^{++}$~\cite{jiang2023mlicpp}} &
\rotatebox[origin=c]{45}{WeConvene~\cite{fu2024weconvene}} &
\rotatebox[origin=c]{45}{LALIC~\cite{feng2025linear}} &
\rotatebox[origin=c]{45}{DCAE~\cite{lu2025learned}} &
\rotatebox[origin=c]{45}{MLICv2} \\
\midrule
Parameters (M) & 50.0 & 41.9 & 75.2 & 100.0 & 75.9 & 69.8 & 83.4 & 83.5 & 105.5 & 63.2 & 119.2 & 84.3 \\
\bottomrule
\end{tabular}}
\label{tab:param}
\end{table*}
\begin{table*}[t]
\footnotesize
\centering
\renewcommand\arraystretch{1}
\caption{Skip Ratio, Encoding Acceleration and Decoding Acceleration at different qualities on Kodak and CLIC Pro Val.}
\setlength{\tabcolsep}{1.8mm}{
\begin{tabular}{@{}c|cccccc|cccccc@{}}
\toprule
& \multicolumn{6}{c|}{Kodak} & \multicolumn{6}{c}{CLIC Pro Val} \\
$\lambda$ & 0.0018 & 0.0035 & 0.0067 & 0.0130 & 0.0250 & 0.0483 & 0.0018 & 0.0035 & 0.0067 & 0.0130 & 0.0250 & 0.0483 \\
\midrule
Skip Ratio (\%)       & 88.32 & 82.73 & 77.09 & 69.76 & 66.44 & 57.26 & 92.95 & 88.20 & 83.72 & 79.72 & 77.28 & 68.81 \\
Enc Acceleration (\%) &  9.98 &  9.48 &  9.24 &  8.81 &  7.31 &  7.22 & 18.48 & 17.12 & 16.88 & 16.62 & 16.15 & 14.21 \\
Dec Acceleration (\%) & 13.61 & 13.21 & 12.15 & 11.97 & 10.18 &  9.62 & 27.52 & 25.85 & 25.37 & 24.21 & 23.36 & 20.99 \\
\bottomrule
\end{tabular}}
\label{tab:skip_combined}
\end{table*} 
\begin{figure*}
  \centering
  \subfloat{
    \includegraphics[scale=0.318]{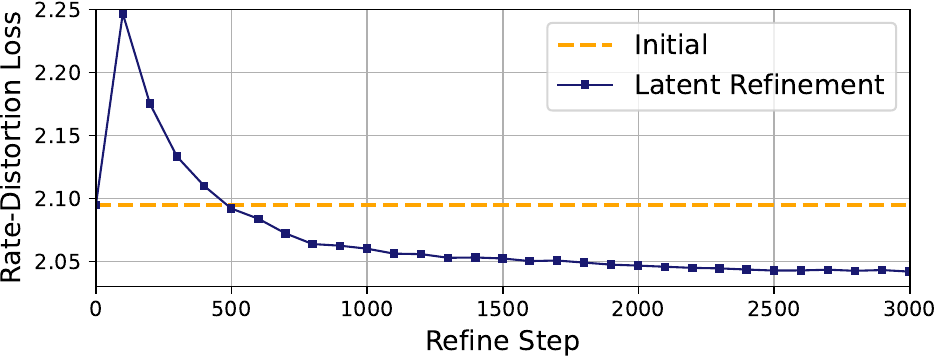}}
    \subfloat{
      \includegraphics[scale=0.318]{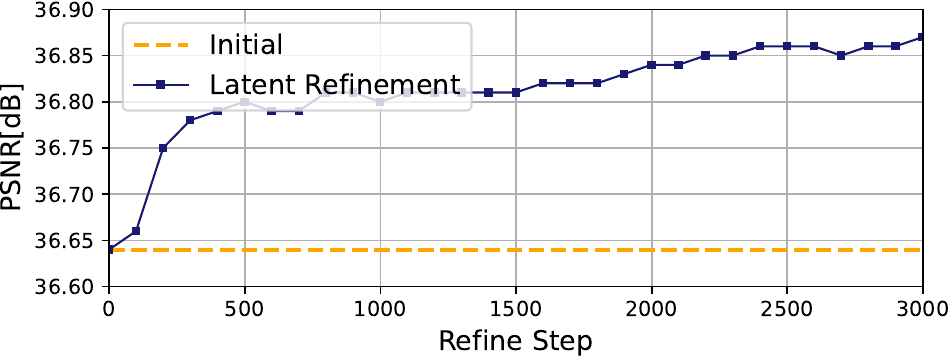}}
  \subfloat{
    \includegraphics[scale=0.318]{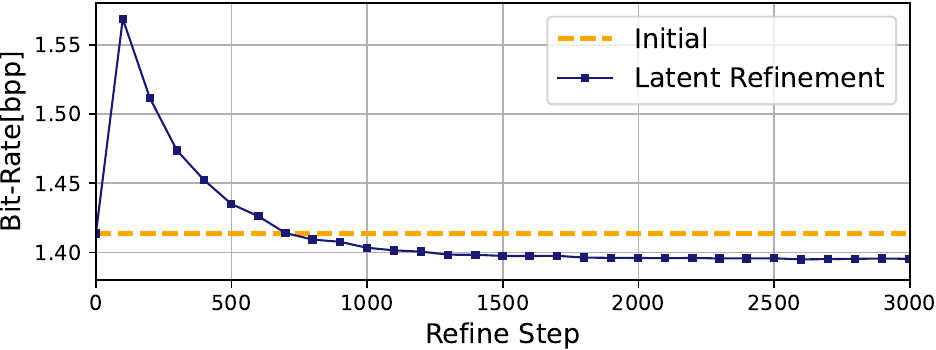}}
  \caption{Changes in loss, PSNR and bpp during the refinement of latent representation and side information of Kodim01 from Kodak dataset~\cite{kodak}.
  ``Initial'' denotes the original loss, psnr, bpp when using the latent representation and side information outputed by MLICv2.}
  \label{fig:refine}
\end{figure*}
{\begin{definition}[Mean Attention Weight]
In the channel reweighting module, the attention map is computed as 
$\boldsymbol{M} = \textrm{Softmax}(\boldsymbol{Q}^\top \boldsymbol{K}) \in \mathbb{R}^{C \times C}$,
where $C$ is the number of channels. 
The attention map is row-wise normalized via the Softmax function:
$
  \boldsymbol{M}_{i,j} = \frac{\exp(\boldsymbol{q}_i^\top \boldsymbol{k}_j)}{\sum_{j^{\ast}=1}^C\exp(\boldsymbol{q}_i^\top \boldsymbol{k}_{j^\ast})},
$ 
where $\boldsymbol{q}_i$ and $\boldsymbol{k}_j$ are the $i$-th and $j$-th channel 
vectors of $\boldsymbol{Q}$ and $\boldsymbol{K}$, respectively.
This normalization ensures that each row of $\boldsymbol{M}$ sums to 1, 
meaning that $\boldsymbol{M}_{i,j}$ reflects the relative influence of input channel $j$ 
on output channel $i$ when computing the weighted sum:
$
  \boldsymbol{o}_i = \sum_{j=1}^C \boldsymbol{M}_{i,j} \boldsymbol{v}_j,
$
where $\boldsymbol{v}_j$ is $j$-th channel 
vector $\boldsymbol{V}$.
The Mean Attention Weight for each input channel $j$ is then defined as the average 
over the $j$-th column of $\boldsymbol{M}$:
$
  \tilde{\boldsymbol{M}}_j = \frac{1}{C}\sum_{i=1}^C \boldsymbol{M}_{i,j},
$
{which reflects the overall influence of input channel $j$ across all output channels.}
\end{definition}}
{As observed, channels dominated 
by noise tend to receive lower Mean Attention Weight, while those containing richer structural information 
or less noise are 
assigned higher weight. This demonstrates that our context 
reweighting module effectively emphasizes more informative contexts, supporting its contribution to the performance gains.}
\par 
To validate the effectiveness of the proposed 2D RoPE, 
the 2D RoPE is replaced by original additive relative position embedding~\cite{liu2021swin}.
The performance degradation caused by the replacement demonstrates the superiority of the proposed 2D RoPE.
{Inspired by some 2D positional encoding designs~\cite{carion2020end, ramachandran2019stand, wang2020axial}, which split the feature tensor along 
the channel dimension and apply positional encoding along the horizontal axis to one half 
and the vertical axis to the other, we also experimented with a similar strategy. 
Specifically, we divided the input tensor along the channel dimension into two 
parts and applied RoPE-based rotations along the horizontal axis and vertical axis, respectively.
We found that this alternative design does not lead to a performance 
difference.
We also conduct experiments on the baseline model using 2D RoPE \textit{without} learnable $\theta_x$ and $\theta_y$.
This variant results in approximately a 0.4\% performance drop.
}\par
The validate the effectiveness of the gate block, the gate block is replaced by 
Multilayer Perceptron (MLP), which leads to performance loss.\par
{Overall, the newly introduced modules in the entropy model yield substantial
performance gains with minimal additional computational cost, 
as demonstrated in our ablation studies and Table~\ref{tab:context_complex}. 
As reported in Table~\ref{tab:context_complex},
the MACs of 2D RoPE are almost negligible. The context reweighting and selective compression modules contribute 
only \bm{$0.68\%$} and \bm{$0.34\%$} to the total MACs on the Kodak dataset, respectively.
This efficiency is largely due to the fact that the proposed modules for entropy modeling operate in the latent space, where the spatial resolution is only $\bm{\frac{1}{16} \times \frac{1}{16}}$ of the input image.
\textit{Such a drastic reduction in resolution greatly reduces the overall computational burden.}}

\subsubsection{Analysis of Iterative Latent Refinement}

In case 1, the latent refinement is removed, which leads to around $4.89\%$ bit-rate increase compared to MLICv2$^+$.
{This significant performance improvement can be attributed to \textit{instance-specific adaptation}. 
While our MLICv2 model is trained on a diverse dataset to achieve good average performance across various images, 
each individual image possesses unique characteristics such as texture complexity, edge distribution, and local redundancy patterns. 
ILR enables the latent representation $\hat{\boldsymbol{y}}$ and side information $\hat{\boldsymbol{z}}$ 
to be specifically optimized for each input image's rate-distortion characteristics, 
effectively bridging the gap between generalized encoder mappings ($g_a$ and $h_a$) and input-specific optimal representations.}
The evolution of loss, PSNR, and bpp during refinement is illustrated in Fig.~\ref{fig:refine}.
{During the initial refinement steps, both bit-rate and PSNR increase simultaneously, 
indicating that ILR adjusts latent values near quantization boundaries to better balance rate and distortion for the specific input. 
The subsequent phase exhibits decreasing bit-rate accompanied by further PSNR improvement, 
demonstrating that the refinement process successfully discovers a more optimal operating point in the rate-distortion space 
that remains inaccessible to the fixed encoder network. This validates the effectiveness of ILR in exploiting the full potential of our architectural framework.}
\section{Conclusion}
\label{sec:conclusion}
In this paper, we present MLICv2 and MLICv2$^+$, enhanced versions of the previous MLIC series that address observed limitations 
in transform design and entropy modeling. To overcome these limitations, we introduce efficient simple token mixing blocks 
that substantially improve representational capacity while maintaining computational efficiency. 
Our MetaFormer-based design achieves comparable performance to complex transformer alternatives with reduced overhead.
For entropy modeling, we propose hyperprior-guided global context prediction, context reweighting mechanisms, 
and novel 2D Rotary Positional Embedding, collectively providing richer contextual information. The guided selective compression further accelerates encoding/decoding through intelligent zero-element skipping.
Additionally, by employing SGA, we demonstrate the potential for performance improvements within our architectural framework.
This validation analysis reveals the potential for further architectural improvements and provides valuable insights for future development directions.
These innovations collectively establish new state-of-the-art performance benchmarks, with BD-Rate reductions of up to 20\% compared to 
VTM-17.0 Intra while maintaining competitive computational complexity. 
Furthermore, to continue exploring the performance upper bounds via encoding-time scaling, we plan to investigate the transmission of update 
parameters~\cite{tsubota2023universal,shen2023dec,lv2023dynamic} to the decoder side.
\par
\textbf{Limitations}:
Although our MLICv2 and MLICv2$^+$ achieve state-of-the-art performance, several limitations warrant future attention. 
First, the performance potential exploration through SGA may lead to {bit-rate} variations~\cite{zhang2024practical}, 
requiring practical online encoder optimization strategies for precise {bit-rate} control. Second, MLICv2/MLICv2$^+$ may encounter 
cross-platform decoding issues~\cite{balle2018integer,he2022post,shi2023rate,yang2025learned}. 
We plan to investigate model quantization techniques~\cite{han2015deep,he2022post,shi2023rate} 
and chain coding-based latent compression~\cite{yang2025learned} to address deployment challenges.

\bibliography{main}
\bibliographystyle{plain}

\end{document}